\documentclass[aps,prd,twocolumn,showpacs,showkeys,nofootinbib]{revtex4}
\pdfoutput=1
\usepackage{epsfig}
\usepackage{color}
\usepackage{amsmath}
\usepackage{graphicx}
 \usepackage{multirow}
\newcommand{\beq}{\begin{equation}}
\newcommand{\eeq}{\end{equation}}
\newcommand{\bqa}{\begin{eqnarray}}
\newcommand{\eqa}{\end{eqnarray}}

\newcommand{\bold}{\textbf}

\def\bfsigma{\mbox{\boldmath $\sigma$}}

\begin{document}

\title{Relativistic corrections to the form factors of $B_c$ into $P$-wave orbitally excited charmonium}
\author{Ruilin Zhu}\email{rlzhu@njnu.edu.cn}
\affiliation{
 Department of Physics and Institute of Theoretical Physics,
Nanjing Normal University, Nanjing, Jiangsu 210023, China\\
 }

\begin{abstract}
We investigated the form factors of the $B_{c}$ meson into $P$-wave
orbitally excited charmonium using the nonrelativistic QCD effective
theory. Through the analytic computation, the next-to-leading order
relativistic corrections to the form factors were obtained, and the
asymptotic expressions were studied in the infinite bottom quark mass
limit. Employing the general form factors, we discussed the exclusive
decays of the $B_{c}$ meson into $P$-wave orbitally excited charmonium
and a light meson. We found that the relativistic corrections lead to
a large correction for the form factors, which
makes the branching ratios of the decay channels $\mathcal{B}(B_{c}
^{\pm }\to \chi _{cJ}(h_c) +\pi ^{\pm }(K^{\pm }))$ larger. These results are
useful for the phenomenological analysis of the $B_{c}$ meson decays
into $P$-wave charmonium, which shall be tested in the LHCb experiments.


\end{abstract}

\maketitle

\section{Introduction}

The $B_{c}$ meson is composed of two different heavy flavors, which has
attracted more attentions from both the theoretical and experimental
aspects in recent years. Because two different heavy quarks are hard to
produce with a strong coupling suppressed factor $\alpha _{s}^{2}$ in the
electro-positron collider, the precisely studies of the $B_{c}$ meson
properties have to refer to hadron-hadron colliders, exemplified by the
discovery of the $B_{c}$ meson in the Tevatron~\cite{Abe:1998wi}.

The cross section of the $B_{c}$ meson at the Large Hadron Collider
(LHC) is expected at the level of dozens of $n
b$~\cite{Chang:2003cr}, so a large number of the $B_{c}$ meson
events can be produced and reconstructed at the LHCb experiments. LHC
provides a solid platform to precisely investigate the production and
decay properties of the $B_{c}$ meson. Through the precise analysis of
the production and decay events, one can obtain more information on the
hadronization mechanism of the heavy quarkonium and $B_{c}$ meson, and
also test the nonrelativistic QCD (NRQCD) effective
theory~\cite{Bodwin:1994jh}.

In literatures, the $B_{c}$ meson spectra are calculated in the
framework of the QCD-motivated relativistic quark model based on the
quasi-potential approach~\cite{Ebert:2011jc}, where many higher
excited states with different spin, orbital and radial quantum numbers
were predicted. An excited state with the mass of $6842\pm 4\pm 5~\mbox{MeV}$
has been observed by the ATLAS Collaboration recently~\cite{Aad:2014laa}, which can be regarded as the possible candidate of
the $B_{c}(2S)$ state.

From a theoretical point of view, the form factors of the $B_{c}$ meson into a
charmonium have been investigated in many works. Some schemes are
employed such as the perturbative QCD (PQCD)
approach~\cite{Du:1988ws,Sun:2008ew,Wen-Fei:2013uea,Rui:2014tpa},
QCD sum rules (QCD
SR)~\cite{Colangelo:1992cx,Kiselev:1993ea,Kiselev:1999sc,Azizi:2009ny},
Light-cone sum rules (LCSR)~\cite{Huang:2007kb}, the relativistic
quark model
(RQM)~\cite{Nobes:2000pm,Ebert:2003cn,Ivanov:2005fd,Ebert:2010zu},
the nonrelativistic constituent quark model
(NCQM)~\cite{Hernandez:2006gt}, the light-front quark model
(LFQM)~\cite{Wang:2008xt,Wang:2009mi,Ke:2013yka}, NRQCD
approach~\cite{Chang:1992pt,Chang:2001vq,Chang:2001pm,Kiselev:2001zb,Bell:2005gw,Qiao:2011yz,Qiao:2012vt,Qiao:2012hp,Shen:2014msa,Zhu:2017lqu}.
It is realized that the decay rates of $B_{c}$ exclusive decays to a
charmonium and a light meson such as $B_{c} \to J/\psi +\pi $ and
$B_{c} \to J/\psi +K$ depend mainly on the form factors at the maximum
momentum recoil point. The careful investigation of the form factors is
important to precisely predict these decay rates. In NRQCD approach, the
relativistic corrections to the form factors of the $B_{c}$ meson into
$S$-wave charmonium have been computed in Ref.~\cite{Zhu:2017lqu},
where the relativistic corrections can bring about additional 15\%--30\%
contributions and thus can not be ignored. However, the relativistic
corrections to the form factors of the $B_{c}$ meson into $P$-wave
charmonium have not been studied yet.

On the other hand, the first evidence for the $B_{c}$ meson decays into
$P$-wave charmonium was observed by the LHCb Collaboration in
2016~\cite{Aaij:2016xas}. Through the analysis of $B_{c}^{+}
\to K^{+} K^{-} \pi ^{+}$ decays with the data corresponding to an
integrated luminosity of $3~\mbox{fb}^{-1}$ collected in $pp$ collisions at
centre-of-mass energies of 7~TeV and 8~TeV, the LHCb Collaboration have
found the evidence for the decay $B_{c}^{+}\to \chi _{c0} \pi ^{+}$ at a
significance of 4 standard deviations. Along with the accumulation of
data at LHC, the experimental studies of the $B_{c}$ meson decays into
$P$-wave charmonium will become accessible. Future measurements in these experiments will definitely improve the degree of accuracy of the
form factors of the $B_{c}$ meson into $P$-wave charmonium.

There are also some works on the relativistic corrections to the light
cone distribution amplitudes of the $B_{c}$
meson~\cite{Xu:2016dgp,Wang:2017bgv}. The recent works on the
$B_{c}$ two-body and quasi-two-body decays can be found in
Refs.~\cite{Chang:2014jca,Li:2016cou,Liu:2017cwl,Ma:2017aie,Mohammadi:2017nml,He:2016xvd}.
In this paper, we will calculate the relativistic corrections to the form
factors of the $B_{c}$ meson decays into $P$-wave charmonia, i.e.
$h_{c}$ and $\chi _{cJ}$ with $J=0,~1,~2$. Then we will employ the form
factors into the phenomenological analysis of the exclusive
$B_{c}$ two-body decays.

The paper is organized as the following. In Sec.~\ref{II}, we will give
the formulae for the form factors of $B_{c}$ into $P$-wave charmonium.
In Sec.~\ref{III}, we will give the analytic expressions of the
relativistic corrections to the form factors of $B_{c}$ into $P$-wave
charmonium. Then the form factors will be investigated in the heavy
bottom quark limit. In Sec.~\ref{IV}, we will give the branching ratios
of $B_{c} \to h_{c} +\pi $, $B_{c} \to h_{c} +K $, $B_{c} \to \chi
_{cJ} +\pi $, and $B_{c} \to \chi _{cJ} +K $. We will summarize and
conclude in the end.

\section{form factors of $B_c$ into $P$-wave charmonium \label{II}}
\subsection{NRQCD effective theory}
The NRQCD Lagrangian is written as~\cite{Bodwin:1994jh}:
\begin{eqnarray}
{\mathcal L}_{\rm NRQCD} &=&
\psi^\dagger \left( i D_t + {{\bf D}^2 \over 2m_Q} \right) \psi
+ {c_F \over 2 m} \psi^\dagger \bfsigma \cdot g_s {\bf B} \psi
\nonumber\\
&+& \psi^\dagger {{\bf D}^4 \over 8m^3_Q} \psi+{c_D\over 8 m^2_Q} \psi^\dagger ({\bf D}\cdot g_s {\bf E}- g_s {\bf E}\cdot {\bf D})\psi
\nonumber\\
&+&{i c_S\over 8 m^2_Q} \psi^\dagger \bfsigma \cdot ({\bf D}\times g_s {\bf E}- g_s {\bf E}\times {\bf D})\psi
\nonumber\\
&+& \left(\psi \rightarrow i \sigma ^2 \chi^*, A_\mu \rightarrow - A_\mu^T\right) +
{\mathcal L}_{\rm light} \,,
\label{NRQCD:Lag}
\end{eqnarray}
where the Lagrangian for the light quarks and gluons is denoted as ${\mathcal L}_{\rm light}$,
which is identical to the related Lagrangian in QCD. $m_Q$ is the heavy quark mass.
The coefficients $c_D$, $c_F$, and $c_S$ can be perturbatively calculated through the matching techniques between NRQCD and QCD.

The NRQCD effective theory has been established for a long time
 by Bodwin, Braaten, and Lepage~\cite{Bodwin:1994jh}, and has successfully achieved
 the factorization formulae for the heavy quarkonium in many production and decay processes.
However, the precise calculation of the short-distance coefficients of the NRQCD long-distance matrix elements (LDMEs) in these processes is still
a challenging topic.

In NRQCD effective theory, the decay width (cross section) of a heavy quarkonium $H$ is generally  factorized as~\cite{Bodwin:1994jh}
\begin{eqnarray}
\Gamma_H=\sum_n\frac{C_n(\mu)}{m_Q^{d_n-3}}\langle H|{\cal O}_n(\mu)|H\rangle\,,\\
\sigma_H=\sum_n\frac{C'_n(\mu)}{m_Q^{d^H_n-2}}\langle 0|{\cal O}^{H}_n(\mu)|0\rangle\,,
\end{eqnarray}
where $\langle H|{\cal O}_n(\mu)|H\rangle$ and $\langle 0|{\cal O}^{H}_n(\mu)|0\rangle$ are the NRQCD LDMEs, which involve nonperturbative information below the scale $m_Q$. The short-distance coefficients
$C_n(\mu)$ include the effects of order $m_Q$ or larger. The NRQCD operators ${\cal O}_n(\mu)$ annihilate
 a heavy quark-antiquark pair, while ${\cal O}^{H}_n(\mu)$ generate a
 a heavy quark-antiquark pair.

The leading order  NRQCD decay operators for  $S$-wave heavy quarkonium can be written as
\begin{eqnarray}
\mathcal{O}(^{1}S_{0}^{[1]})&=&\psi^{\dagger}\chi\chi^{\dagger}\psi,\\
\mathcal{O}(^{3}S_{0}^{[1]})&=&\psi^{\dagger}\bfsigma\chi\cdot\chi^{\dagger}\bfsigma\psi.
\end{eqnarray}

The leading order  NRQCD production operators for $P$-wave quarkonium  are
\begin{eqnarray}
\mathcal{O}^H(^{1}P_{1}^{[1]})&=&\chi^{\dagger}(-\frac{i}{2}{\overleftrightarrow{ {D}^i}})\psi(a_H^\dagger a_H)\psi^{\dagger}(-\frac{i}{2}{\overleftrightarrow{ { D}^i}})\chi,\\
\mathcal{O}^H(^{3}P_{0}^{[1]})&=&\frac{1}{3}\chi^{\dagger}(-\frac{i}{2}{\overleftrightarrow{ {\bold D}}}\cdot \bfsigma)\psi(a_H^\dagger a_H)\psi^{\dagger}(-\frac{i}{2}{\overleftrightarrow{ {\bold D}}}\cdot \bfsigma)\chi,\nonumber\\\\
\mathcal{O}^H(^{3}P_{1}^{[1]})&=&\frac{1}{2}\chi^{\dagger}(-\frac{i}{2}{\overleftrightarrow{ {\bold D}}}\times \bfsigma)\psi(a_H^\dagger a_H)\nonumber\\&&\times\psi^{\dagger}(-\frac{i}{2}{\overleftrightarrow{ {\bold D}}}\times \bfsigma)\chi,\\
\mathcal{O}^H(^{3}P_{2}^{[1]})&=&\chi^{\dagger}(-\frac{i}{2}{\overleftrightarrow{ {D}^i}}\sigma^j)\psi(a_H^\dagger a_H)\psi^{\dagger}(-\frac{i}{2}{\overleftrightarrow{ { D}^i}} \sigma^j)\chi.
\end{eqnarray}

The next-to-leading order  relativistic correction decay operators for $S$-wave heavy quarkonium  are
\begin{eqnarray}
\mathcal{P}(^{1}S_{0}^{[1]})&=&\frac{1}{2}\left[\psi^{\dagger}\chi\cdot\chi^{\dagger}(-\frac{i}{2}{\overleftrightarrow{ {\bold D}}})^2\psi+h.c.\right],\\
\mathcal{P}(^{3}S_{1}^{[1]})&=&\frac{1}{2}\left[\psi^{\dagger}\bfsigma
\chi\cdot\chi^{\dagger}\bfsigma(-\frac{i}{2}{\overleftrightarrow{ {\bold D}}})^2\psi+h.c.\right],
\end{eqnarray}
where the $h.c.$ denotes the related complex conjugate term.

The next-to-leading order  relativistic correction production operators for $P$-wave heavy quarkonium  are
\begin{eqnarray}
\mathcal{P}^H(^{1}P_{1}^{[1]})&=&\frac{1}{2}\left[\chi^{\dagger}(-\frac{i}{2}{\overleftrightarrow{ {D}^i}})\psi(a_H^\dagger a_H)\psi^{\dagger}(-\frac{i}{2}{\overleftrightarrow{ { D}^i}})\right.\nonumber\\&&\left.\times(-\frac{i}{2}{\overleftrightarrow{ {\bold D}}})^2\chi+h.c.\right],\\
\mathcal{P}^H(^{3}P_{0}^{[1]})&=&\frac{1}{2}\left[\frac{1}{3}\chi^{\dagger}(-\frac{i}{2}{\overleftrightarrow{ {\bold D}}}\cdot \bfsigma)\psi(a_H^\dagger a_H)\psi^{\dagger}(-\frac{i}{2}{\overleftrightarrow{ {\bold D}}}\cdot \bfsigma)\right.\nonumber\\&&\left.\times(-\frac{i}{2}{\overleftrightarrow{ {\bold D}}})^2\chi+h.c.\right],\\
\mathcal{P}^H(^{3}P_{1}^{[1]})&=&\frac{1}{2}\left[\frac{1}{2}\chi^{\dagger}(-\frac{i}{2}{\overleftrightarrow{ {\bold D}}}\times \bfsigma)\psi(a_H^\dagger a_H)\right.\nonumber\\&&\left.\times\psi^{\dagger}(-\frac{i}{2}{\overleftrightarrow{ {\bold D}}}\times \bfsigma)(-\frac{i}{2}{\overleftrightarrow{ {\bold D}}})^2\chi+h.c.\right],\\
\mathcal{P}^H(^{3}P_{2}^{[1]})&=&\frac{1}{2}\left[\chi^{\dagger}(-\frac{i}{2}{\overleftrightarrow{ {D}^i}}\sigma^j)\psi(a_H^\dagger a_H)\psi^{\dagger}(-\frac{i}{2}{\overleftrightarrow{ { D}^i}} \sigma^j)\right.\nonumber\\&&\left.\times(-\frac{i}{2}{\overleftrightarrow{ {\bold D}}})^2\chi+h.c.\right].
\end{eqnarray}

\subsection{Covariant projection method}
The $B_c$ meson has two different heavy flavors, thus we introduce two heavy quarks $Q$ and $Q'$. And the heavy quarkonium can be described when $Q'=Q$. Let $p_1$ and $p_2$ represent the momenta for the heavy quark $Q$ and anti-quark $\bar{Q^\prime}$, respectively. The momenta can be decomposed as
\begin{eqnarray}
p_1 &=&  \alpha \,P_{H}-k,\\\
p_2 &=&  \beta\, P_{H}+k,
\end{eqnarray}
where $P_{H}=p_1+p_2$ being the heavy quarkonium momentum, $k$ being a half of the relative momentum between the
quark pair, and $\alpha+\beta=1$.

In the rest frame of the  heavy quarkonium $H$, the explicit expressions are
\begin{eqnarray}
P_{H}^\mu &=&  (E_1+E_2,0),\\
k^\mu &=&  (0,\bold{k}\,),\\
p_1^\mu &=&  (E_1,-\bold{k}\,),\\
p_2^\mu &=&  (E_2,\bold{k}\,),
\end{eqnarray}
 where the heavy quark on-shell conditions being $E_1=\sqrt{m_1^2-k^2}$, $E_2=\sqrt{m_2^2-k^2}$ and $k^2=-\bold{k}^2$ with the heavy quark masses $m_1$ and $m_2$, $\alpha={\sqrt{m_1^2-k^2}}/({\sqrt{m_1^2-k^2}+\sqrt{m_2^2-k^2}})$,  and $\beta=1-\alpha$.

The Dirac spinors for the heavy quark $Q$ and anti-quark $\bar{Q^\prime}$ are
\begin{eqnarray}
u_1(p_1,\lambda) &=&  \sqrt{\frac{E_1+m_1}{2E_1}}\left(
                                           \begin{array}{ll}
                             ~~~~\xi_\lambda \\
\frac{\vec{\sigma}\cdot \overrightarrow{p_1}}{E_1+m_1}\xi_\lambda
                                           \end{array}
                                         \right)\,,
\end{eqnarray}

\begin{eqnarray}
v_2(p_2,\lambda) &=&  \sqrt{\frac{E_2+m_2}{2E_2}}\left(
                                           \begin{array}{ll}
\frac{\vec{\sigma}\cdot \overrightarrow{p_2}}{E_2+m_2}\xi_\lambda\\
                             ~~~~\xi_\lambda\end{array}
                                         \right)\,.
\end{eqnarray}

 The short-distance coefficients of  NRQCD LDMEs will be obtained in principle through the matching techniques. For a certain process, one usually extract the short-distance coefficients using the covariant projection method. The corresponding projection operators are
\begin{widetext}
\begin{eqnarray}
\Pi_{S=0,1}(k) &=&  -i\sum_{\lambda_1,\lambda_2} u_b(p_1,\lambda_1)\bar{v}_c(p_2,\lambda_2)\langle\frac{1}{2}\lambda_1\frac{1}{2}\lambda_2|S S_z\rangle\otimes \frac{\bold{1}_c}{\sqrt{N_c}}\nonumber\\
&=&\frac{i}{4\sqrt{2 E_1 E_2}\omega}(\alpha \,p\!\!\!\slash_{H}-k\!\!\!\slash+m_1)\frac{p\!\!\!\slash_{H}+E_1+E_2}{E_1+E_2}\Gamma_S
(\beta\,p\!\!\!\slash_{H}+k\!\!\!\slash-m_2)\otimes \frac{\bold{1}_c}{\sqrt{N_c}}\,,\label{projection}
\end{eqnarray}
\end{widetext}
where $\omega=\sqrt{E_1+m_1}\sqrt{E_2+m_2}$.  For the spin-singlet combination, we have the spin $S=0$ and $\Gamma_{S=0}=\gamma^5$. For the spin-triplet combination, we have the spin $S=1$ and $\Gamma_{S=1}=\varepsilon\!\!\!\slash^*_{H}=\varepsilon^*_\mu(p_H) \gamma^\mu$.

In the nonrelativistic  bound-state picture, the heavy quarkonium can be described by the nonrelativistic wave function. For $P$-wave charmonium, one has to expand the amplitude to the first order in $k$. Thus the amplitude can be obtained by \cite{Song:2003yc}
\begin{eqnarray}
&&{\cal M}(B_c\to \;^{2S+1}P_J(c\bar{c})+X)\nonumber\\&=&
 \sum_{L_z,S_z}\langle 1L_z;SS_z|JJ_z\rangle\int\frac{d^4k}{(2\pi)^3}k^\alpha\delta(k^0-\frac{|\vec{k}|^2}{m_{H}})\psi^*(k)_H
 \nonumber\\ &&\times \mathrm{Tr}\left[{\cal O}_\alpha (0)\Pi_{S}(0)+{\cal O} (0)\Pi_{S,\alpha}(0)\right],
\end{eqnarray}
where
\begin{eqnarray}
{\cal O}_\alpha (0)=\frac{\partial {\cal O}(k)}{\partial k^\alpha}\mid_{k=0},\quad  \Pi_{S,\alpha}(0)=\frac{\partial \Pi_{S}(k)}{\partial k^\alpha}\mid_{k=0}.&&
\end{eqnarray}

After integrating the zero-component $k^0$ of the momentum, the integral is proportional to the derivative of the radial wave function at the origin
\begin{eqnarray}
\int\frac{d^3k}{(2\pi)^3}k^\alpha \psi^*(k)_H =i\varepsilon^{*\alpha}(L_z)\sqrt{\frac{3}{4\pi}}{\cal R}'(0)_H,&&
\end{eqnarray}
where the derivative of the radial wave function at the origin can be related to
the derivative of wave functions at the origin of $P$-wave charmonia  by $\psi'(0)_{H }=\sqrt{\frac{1}{4\pi}}{\cal R}'(0)_H$.

\subsection{Form factors}
$P$-wave orbitally excited charmonia include the spin-singlet $h_c(^1P_1)$, the scalar $\chi_{c0}(^3P_0)$,
the axial-vector meson $\chi_{c1}(^3P_1)$, and the tensor meson $\chi_{c2}(^3P_2)$. Thus the form factors of the $B_c$ meson into $P$-wave orbitally excited charmonium can be defined accordingly.

The form factors of the $B_c$ meson into $h_c(^1P_1)$ are defined as
\begin{eqnarray}
&&\langle h_{c}(p,\varepsilon^{*})\vert \bar c \gamma^{\mu}b\vert B_{c}(P)\rangle
=-i[2
m_{h_c}A^{h_c}_{0}(q^{2})\frac{\varepsilon^{*}\cdot q}{q^{2}}q^{\mu}
\nonumber\\&&~~~~~-A^{h_c}_{2}(q^{2})\frac{\varepsilon^{*}\cdot
q}{m_{B_{c}}+m_{h_c}}(
P^{\mu}+p^{\mu}-\frac{m_{B_{c}}^{2}-m_{h_c}^{2}}{q^{2}}q^{\mu})
\nonumber\\&&~~~~~+(m_{B_{c}}+m_{h_c})A^{h_c}_{1}(q^{2})
(\varepsilon^{*\mu}-\frac{\varepsilon^{*}\cdot q}{q^{2}} q^{\mu})]\,,
\end{eqnarray}
\begin{eqnarray}
 && \langle h_c(p,\varepsilon^{*})\vert \bar c \gamma^{\mu}\gamma^{5}b\vert
B_{c}(P)\rangle =\frac{2
V^{h_c}(q^{2})}{m_{B_{c}}+m_{h_c}}\epsilon^{\mu\nu\rho\sigma}
\varepsilon_{\nu}^{*}p_{\rho}P_{\sigma}\,,\nonumber\\
\end{eqnarray}
where the momentum transfer is defined as $q=P-p$ with the $B_c$ meson momentum $P$ and the final charmonium momentum $p$. The polarization vector or tensor of the final charmonium  is denoted as $\varepsilon^{*}$.
There are only two form factors for the $B_c$ meson into the scalar meson, which are defined as
\begin{eqnarray}
&&\langle \chi_{c0}(p)\vert \bar c \gamma^{\mu}\gamma^{5}b\vert B_{c}(P)\rangle
=[f^{\chi_{c0}}_{0}(q^{2})
\frac{m_{B_{c}}^{2}-m_{\chi_{c0}}^{2}}{q^{2}}q^{\mu}\nonumber\\ &&~~~~+f^{\chi_{c0}}_{+}(q^{2})(P^{\mu}+p^{\mu}-\frac{m_{B_{c}}^{2}-
m_{\chi_{c0}}^{2}}{q^{2}}q^{\mu})](-i)\,.
\end{eqnarray}
 The form factors for the $B_c$ meson into the axial-vector meson can be written as
\begin{eqnarray}
&&\langle \chi_{c1}(p,\varepsilon^{*})\vert \bar c \gamma^{\mu}b\vert B_{c}(P)\rangle
=-i[2
m_{\chi_{c1}}A^{\chi_{c1}}_{0}(q^{2})\frac{\varepsilon^{*}\cdot q}{q^{2}}q^{\mu}
\nonumber\\&&~~~-A^{\chi_{c1}}_{2}(q^{2})\frac{\varepsilon^{*}\cdot
q}{m_{B_{c}}+m_{\chi_{c1}}}(
P^{\mu}+p^{\mu}-\frac{m_{B_{c}}^{2}-m_{\chi_{c1}}^{2}}{q^{2}}q^{\mu})
\nonumber\\&&~~~+(m_{B_{c}}+m_{\chi_{c1}})A^{\chi_{c1}}_{1}(q^{2})
(\varepsilon^{*\mu}-\frac{\varepsilon^{*}\cdot q}{q^{2}} q^{\mu})]\,,
\end{eqnarray}
\begin{eqnarray}
 && \langle \chi_{c1}(p,\varepsilon^{*})\vert \bar c \gamma^{\mu}\gamma^{5}b\vert
B_{c}(P)\rangle =\frac{2
V^{\chi_{c1}}(q^{2})}{m_{B_{c}}+m_{\chi_{c1}}}\epsilon^{\mu\nu\rho\sigma}
\varepsilon_{\nu}^{*}p_{\rho}P_{\sigma}.\nonumber\\
\end{eqnarray}
The form factors for the $B_c$ meson into the tensor meson can be written as
\begin{eqnarray}
&&\langle \chi_{c2}(p,\varepsilon^{*})\vert \bar c \gamma^{\mu}\gamma^{5}b\vert B_{c}(P)\rangle
=[2
m_{\chi_{c2}}A^{\chi_{c2}}_{0}(q^{2})\frac{\varepsilon^{*\alpha\beta}q_\beta}{q^{2}}q^{\mu}
\nonumber\\&&~-A^{\chi_{c2}}_{2}(q^{2})\frac{\varepsilon^{*\alpha\beta}
q_\beta}{m_{B_{c}}+m_{\chi_{c2}}}(
P^{\mu}+p^{\mu}-\frac{m_{B_{c}}^{2}-m_{\chi_{c2}}^{2}}{q^{2}}q^{\mu})
\nonumber\\&&~+(m_{B_{c}}+m_{\chi_{c2}})A^{\chi_{c2}}_{1}(q^{2})
(\varepsilon^{*\mu\alpha}-\frac{\varepsilon^{*\alpha\beta} q_\beta}{q^{2}} q^{\mu})]\frac{-i P_{\alpha}}{m_{B_c}}\,,\nonumber\\
\end{eqnarray}
\begin{eqnarray}
 && \langle \chi_{c2}(p,\varepsilon^{*})\vert \bar c \gamma^{\mu}b\vert
B_{c}(P)\rangle \nonumber\\=&&\frac{2
V^{\chi_{c2}}(q^{2})}{m_{B_c}(m_{B_{c}}+m_{\chi_{c2}})}\epsilon^{\mu\nu\rho\sigma}
\varepsilon_{\nu\alpha}^{*}p_{\rho}P_{\sigma}P_{\alpha}\,.
\end{eqnarray}

The above definitions of the form factors bring about two benefits.
On the one hand, the decay amplitudes can be described according to these form factors and then
become simple. On the other hand, the form factors in the above formulae are dimensionless parameters,
 which have more wide utilizations in different processes.

The polarization summation for the $^3P_J$ states are
\begin{eqnarray}
&&\sum\limits_{L_z,S_z}\varepsilon^{*\alpha}(L_z)\varepsilon^{*\beta}(S_z)\langle 1L_z;1S_z|00\rangle
\nonumber\\&& =\frac{1}{\sqrt{3}}\left(-g^{\alpha\beta}+\frac{p_H^\alpha p_H^\beta}{m_H^2}\right),
\\
&&\sum\limits_{L_z,S_z}\varepsilon^{*\alpha}(L_z)\varepsilon^{*\beta}(S_z)\langle 1L_z;1S_z|1J_z\rangle
 \nonumber\\&& =\frac{-i\epsilon^{\alpha\beta\gamma\delta}p_{H,\delta}\varepsilon^*_\gamma(p_H,J_z)}{\sqrt{2}m_H},
 \\
 &&\sum\limits_{L_z,S_z}\varepsilon^{*\alpha}(L_z)\varepsilon^{*\beta}(S_z)\langle 1L_z;1S_z|2J_z\rangle
 \nonumber\\&& =\varepsilon^{*\alpha\beta}(p_H,J_z),
\end{eqnarray}
where $\varepsilon^*_\gamma(p_H,J_z)$ is the spin-1 polarization vector with $\varepsilon^*_\gamma(p_H,J_z)p^\gamma_H=0$ and $\varepsilon^{*\alpha\beta}(p_H,J_z)$ is the polarization tensor for the spin-2 system with $\varepsilon^{*\alpha\beta}(p_H,J_z)p_{H,\beta}=0$  which
is symmetric under the exchange $\alpha\leftrightarrow \beta$.

 The five components ($J_z=0,\pm1,\pm2$) of the spin-2 polarization tensor for the $\chi_{c2}(^3P_2)$ state can be constructed via the spin-1 polarization vector as
\begin{eqnarray}
&&\varepsilon^{\alpha\beta}(\pm2)=\varepsilon^{\alpha}(\pm)\varepsilon^{\beta}(\pm),
\nonumber\\&& \varepsilon^{\alpha\beta}(\pm1)=\frac{1}{\sqrt{2}}[\varepsilon^{\alpha}(\pm)\varepsilon^{\beta}(0)
+\varepsilon^{\alpha}(0)\varepsilon^{\beta}(\pm)],
\nonumber\\&& \varepsilon^{\alpha\beta}(0)=\frac{1}{\sqrt{6}}[\varepsilon^{\alpha}(+)\varepsilon^{\beta}(-)
+\varepsilon^{\alpha}(-)\varepsilon^{\beta}(+)]\nonumber
\\&&\quad\quad\quad\quad+\sqrt{\frac{2}{3}}\varepsilon^{\alpha}(0)\varepsilon^{\beta}(0),
\end{eqnarray}
where the explicit structures of the $\varepsilon$ in the initial hadron rest frame can be written as
\begin{eqnarray}
&&\varepsilon^{\alpha}(\pm)=\frac{1}{\sqrt{2}}(0,\mp1,-i,0),
\nonumber\\&& \varepsilon^{\alpha}(0)=\frac{1}{m_{\chi_{c2}}}(|\bold{p}_{\chi_{c2}}|,0,0,p_{\chi_{c2}}^0).
\end{eqnarray}

The sum over polarization for
a spin-1 system is defined as $\Theta^{\alpha\beta}$
\begin{eqnarray}
\sum\limits_{J_z}\varepsilon^{\alpha}(p_H,J_z)\varepsilon^{*\beta}(p_H,J_z)&=&\Theta^{\alpha\beta},
\end{eqnarray}
with
\begin{eqnarray}
\Theta^{\alpha\beta}&=&-g^{\alpha\beta}+\frac{p_H^\alpha p_H^\beta}{m_H^2}.
\end{eqnarray}
The sum over polarization for
a spin-2 system is
\begin{eqnarray}
&&\sum\limits_{J_z}\varepsilon^{\alpha\beta}(p_H,J_z)\varepsilon^{*\alpha'\beta'}(p_H,J_z)\nonumber\\
&&=\frac{1}{2}\left(\Theta^{\alpha\alpha'}\Theta^{\beta\beta'}+\Theta^{\alpha\beta'}
\Theta^{\beta\alpha'}\right)-\frac{1}{3}\Theta^{\alpha\beta}\Theta^{\alpha'\beta'}.
\end{eqnarray}

The  Taylor expansion of the amplitudes in powers of $k^\mu$ is adopted in order to calculate the relativistic corrections to the form factors
\begin{eqnarray}
{\cal M}(k)&=& {\cal M}(0)+\frac{\partial {\cal M}(k)}{\partial k^\mu}\mid_{k=0}k^\mu\nonumber\\
&&+\frac{1}{2!}\frac{\partial^2 {\cal M}(k)}{\partial k^\mu\partial k^\nu}\mid_{k=0}k^\mu k^\nu
\nonumber\\
&&+\frac{1}{3!}\frac{\partial^3 {\cal M}(k)}{\partial k^\mu\partial k^\nu \partial k^\rho}\mid_{k=0}k^\mu k^\nu k^\rho+\ldots.
\end{eqnarray}

\begin{figure}[th]
\begin{center}
\includegraphics[width=0.45\textwidth]{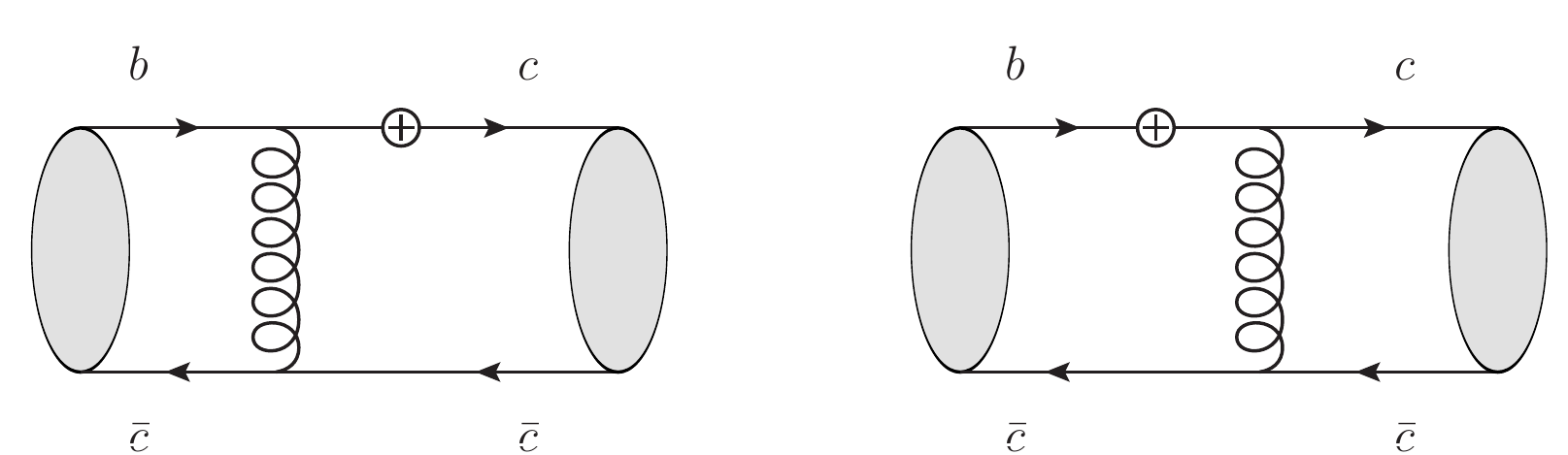}
\end{center}
    \vskip -0.7cm \caption{Feynman diagrams for the form factors of the $B_c$ meson decays into $P$-wave charmonium.}\label{Fig-formfactors}
\end{figure}
According to the Feynman diagrams calculation in Fig.~\ref{Fig-formfactors}, the leading order (LO)  results at ${\cal O}(\alpha_s v^0)$ of the form factors can
be obtained. They are
\begin{widetext}
 \bqa V^{h_c}(q^{2})|_{LO}=\frac{8 \sqrt{2} \pi  (z+1)^{3/2} (3 z+1)  C_F \alpha _s\psi(0)_{B_c}\psi'(0)_{h_c} }{z^{5/2} m_b^4  (y-z+1)^2 (y+z-1)^2},
 \eqa
\bqa A_0^{h_c}(q^2)|_{LO}=-\frac{16 \sqrt{2} \pi \left(z^2-1\right) \left(-y^2 (3 z+2)+5 z^3+8 z^2+9 z+2\right) C_F \alpha _s  \psi(0)_{B_c}\psi'(0)_{h_c}}{z^2 \sqrt{z (z+1)}
   m_b^4 \left((z-1)^2-y^2\right)^3} ,
 \eqa
\bqa A_1^{h_c}(q^2)|_{LO}&=&\frac{8 \sqrt{2} \pi   (z+1)^{3/2}  \left(-y^2+5 z^2+2 z+1\right)C_F \alpha _s \psi(0)_{B_c}\psi'(0)_{h_c}}{z^{5/2}
   m_b^4 \left(y^2-(z-1)^2\right)^2 \left(3 z +1\right)},~~
 \eqa
\bqa A_2^{h_c}(q^2)|_{LO}&=&-\frac{8 \sqrt{2} \pi  \sqrt{z+1} (3 z+1) \left(y^2 (1-3
   z)+15 z^3+17 z^2+17 z-1\right)  C_F \alpha _s\psi(0)_{B_c}\psi'(0)_{h_c}}{z^{5/2} m_b^4 \left((z-1)^2-y^2\right)^3} ,
 \eqa

\bqa f_+^{\chi_{c0}}(q^{2})|_{LO}&=&-\frac{8 \sqrt{\frac{2}{3}} \pi  \sqrt{z (z+1)} \left(-y^4+2 y^2 \left(-2 z^2+z+5\right)+9 z^4+6 z^3-6 z-9\right)  C_F \alpha _s\psi(0)_{B_c}\psi'(0)_{\chi_{c0}}}{z^3
   m_b^4  \left((z-1)^2-y^2\right)^3} , \eqa
\bqa f_0^{\chi_{c0}}(q^{2})|_{LO}&=&-\frac{8 \sqrt{6} \pi   (z (z+1))^{3/2} \left(-y^2 (5 z+3)+9 z^3+9 z^2+11 z+3\right) C_F \alpha _s \psi(0)_{B_c}\psi'(0)_{\chi_{c0}}}{z^4 \left(3 z^2-2 z-1\right)
   m_b^4  \left(y^2-(z-1)^2\right)^2} ,~~
 \eqa

 \bqa V^{\chi_{c1}}(q^{2})|_{LO}=\frac{8 \pi    \sqrt{z+1} (3 z+1)  \left(-y^2 (7 z+5)+11 z^3+15 z^2+17 z+5\right)C_F \alpha _s \psi(0)_{B_c}\psi'(0)_{\chi_{c1}}}{z^{5/2} m_b^4  (-y+z-1)^3
   (y+z-1)^3} ,
 \eqa
\bqa A_0^{\chi_{c1}}(q^2)|_{LO}=-\frac{16 \pi   (z-1) (z+1)^{3/2} C_F \alpha _s\psi(0)_{B_c}\psi'(0)_{\chi_{c1}}}{z^{5/2} m_b^4 \left(y^2-(z-1)^2\right)^2} ,
 \eqa
\bqa A_1^{\chi_{c1}}(q^2)|_{LO}&=&\frac{8 \pi  \sqrt{z+1} \left(-y^2 (9 z+5)+25 z^3+35 z^2+31 z+5\right)C_F \alpha _s \psi(0)_{B_c}\psi'(0)_{\chi_{c1}} }{z^{5/2} m_b^4 \left(y^2-(z-1)^2\right)^2 \left(3 z
   +1\right)} ,~~
 \eqa
\bqa A_2^{\chi_{c1}}(q^2)|_{LO}&=&-\frac{8 \pi  \sqrt{z+1} (3 z+1)   \left(y^2 (z+5)+11 z^3-3 z^2-3 z-5\right)C_F \alpha _s\psi(0)_{B_c}\psi'(0)_{\chi_{c1}}}{z^{5/2} m_b^4
   \left((z-1)^2-y^2\right)^3} ,
 \eqa

\bqa V^{\chi_{c2}}(q^{2})|_{LO}=\frac{96 \sqrt{2} \pi  \sqrt{z+1} (3 z+1)^3 C_F \alpha _s \psi(0)_{B_c}\psi'(0)_{\chi_{c2}}}{z^{3/2}\left((1-z)^2-y^2\right)^3 m_b^4  } ,
 \eqa
\bqa A_0^{\chi_{c2}}(q^2)|_{LO}=\frac{96 \sqrt{2} \pi   (z+1)^{7/2} C_F \alpha _s\psi(0)_{B_c}\psi'(0)_{\chi_{c2}}}{z^{3/2} m_b^4 \left((z-1)^2-y^2\right)^3} ,
 \eqa
\bqa A_1^{\chi_{c2}}(q^2)|_{LO}&=&\frac{32 \sqrt{2} \pi
  \left(\frac{1}{z}+1\right)^{3/2} \left(-y^2 (5 z+3)+9 z^3+17 z^2+19 z+3\right)C_F \alpha _s  \psi(0)_{B_c}\psi'(0)_{\chi_{c2}}}{m_b^4 \left((z-1)^2-y^2\right)^3
   \left(3 z +1\right)} ,~~
 \eqa
\bqa A_2^{\chi_{c2}}(q^2)|_{LO}&=&\frac{32 \sqrt{2} \pi   \left(\frac{1}{z}+1\right)^{3/2} (z+3) (3 z+1)  C_F \alpha _s\psi(0)_{B_c}\psi'(0)_{\chi_{c2}}}{m_b^4
   \left((z-1)^2-y^2\right)^3} ,
 \eqa
 \end{widetext}
where the heavy quark ratio $z$ is defined as $z= m_c/m_b$, and the parameter $y$ is related to the recoil momentum fraction with $y= \sqrt{q^2/m_b^2}$.
Note that we adopt the vacuum-saturation approximation,  the  NRQCD LDMEs can be estimated as $\langle H|\mathcal{O}_n| H\rangle\simeq \langle H|\psi^{\dagger}\mathcal{K}^\prime_n\chi|0\rangle\langle 0|\chi^{\dagger}\mathcal{K}_n\psi| H\rangle$ with $\mathcal{O}_n=\psi^{\dagger}\mathcal{K}^\prime_n\chi\chi^{\dagger}\mathcal{K}_n\psi$.
On the other hand, the  vacuum expectations of the production operators $\mathcal{O}^H_n$ can be estimated as
$\langle 0|\mathcal{O}^H_n| 0\rangle=\langle 0|\chi^{\dagger}\mathcal{K}^\prime_n\psi\sum_m(|H_J,m\rangle\langle H_J,m|)\psi^{\dagger}\mathcal{K}_n\chi| 0\rangle\simeq (2J+1)\langle H|\mathcal{O}_n| H\rangle$ with the angular momentum $J$ of heavy quarkonium. The wave function at the origin of the $B_c$ meson is defined as~\cite{Zhu:2017lqu}
\bqa
\psi(0)_{B_c}&=&\frac{1}{\sqrt{2N_c}}\langle 0|
\chi_b^{\dagger}\psi_c|B_c\rangle.
 \eqa

 The derivative of wave functions at the origin of $P$-wave charmonia  are related to the nonperturbative matrix elements~\cite{Bodwin:1994jh}
\bqa
\psi'(0)_{h_c }\varepsilon^{*i}&=&\frac{1}{\sqrt{2N_c}}\langle h_c(\varepsilon^*)|
\psi^{\dagger}(-\frac{i}{2}{\overleftrightarrow{ { D}^i}})\chi|0\rangle.\\
\psi'(0)_{\chi_{c0} }&=&\frac{1}{\sqrt{3}}\frac{1}{\sqrt{2N_c}}\langle \chi_{c0}|
\psi^{\dagger}(-\frac{i}{2}{\overleftrightarrow{ {\bold D}}}\cdot \bfsigma)\chi|0\rangle.
\eqa
\bqa
\psi'(0)_{\chi_{c1} }\varepsilon^{*i} &=&\frac{1}{\sqrt{2}}\frac{1}{\sqrt{2N_c}}\langle \chi_{c1}(\varepsilon^*)|
\psi^{\dagger}(-\frac{i}{2}{\overleftrightarrow{ {\bold D}}}\times \bfsigma)^i\chi|0\rangle.\nonumber\\\\
\psi'(0)_{\chi_{c2} }\varepsilon^{*ij} &=&\frac{1}{\sqrt{2N_c}}\langle \chi_{c2}(\varepsilon^*)|
\psi^{\dagger}(-\frac{i}{2}{\overleftrightarrow{ { D}^i}} \sigma^j)\chi|0\rangle.
 \eqa

\vskip 0.1 in
\section{Relativistic corrections to the form factors of $B_c$ into $P$-wave charmonium \label{III}}

In this section, we will calculate the relativistic corrections to the form factors. Since both the $B_c$ meson and  the final charmonium  are  bound states composed of two heavy quarks, the relativistic corrections will include two parts: one from the corrections ordered by the charm quark relative velocity in the charmonium; the other from the corrections ordered by the heavy quark relative velocity in the $B_c$ meson.

 In the following, we will define a half of the reduced heavy quark relative momentum as $k=m_{red}v=m_b m_c v/(m_b+m_c)$ inside the $B_c$ meson, and the masses of the $B_c$ meson can be written as
$m_{B_c}=\sqrt{m_c^2-k^2}+\sqrt{m_b^2-k^2}$.  For $P$-wave charmonium, we will define  a half of the charm quark relative momentum as $k'= m_c v'/2$ inside the charmonium, and the masses of the $P$-wave charmonium can be written as
$m_{h_c}\simeq m_{\chi_{c0}}\simeq m_{\chi_{c1}}\simeq m_{\chi_{c2}}=2\sqrt{m_c^2-k'^2}$. Note that the mass corrections of the $B_c$ meson and the charmonium will contribute to the relativistic corrections of the form factors.

\subsection{Calculation steps}

For clarity, we will divide the relativistic corrections into two steps. Firstly, we define the new form factors which do not depend explicitly on the masses of the heavy quarkonium and the $B_c$ meson  as follows~\cite{Qiao:2011yz}
\begin{eqnarray}
&&\langle H(p,\varepsilon^{*})\vert \bar c \gamma^{\mu}b\vert B_{c}(P)\rangle\nonumber\\
&&~~~~~~=-i[a^H_0\varepsilon^{*\mu}+a^H_+\varepsilon^{*}\cdot q p^{\mu}+a^H_-\varepsilon^{*}\cdot q P^{\mu}]
\,,\\
 && \langle H(p,\varepsilon^{*})\vert\bar c \gamma^{\mu}\gamma^{5}b\vert
B_{c}(P)\rangle =g^H\epsilon^{\mu\nu\rho\sigma}
\varepsilon_{\nu}^{*}p_{\rho}P_{\sigma}\,.
\end{eqnarray}
The above formulae can be directly employed for the $h_c$ and $\chi_{c1}$. For the tensor $\chi_{c2}$, the related formulae can be obtained by the replacements $\varepsilon_{\nu}^{*}\to \varepsilon_{\nu\beta}^{*}(\chi_{c2})P_\beta$, $\varepsilon^{*}\cdot q \to \varepsilon^{\alpha\beta*}(\chi_{c2}) q_\alpha  P_\beta$, and $\bar c \gamma^{\mu}b \leftrightarrow\bar c \gamma^{\mu}\gamma^{5}b$.  For the scalar $\chi_{c0}$, the new form factors are defined as
\begin{eqnarray}
&&\langle \chi_{c0}(p)\vert \bar c \gamma^{\mu}\gamma^{5}b\vert B_{c}(P)\rangle\nonumber\\
&&~~~~~~=-i[f^{\chi_{c0}}_p (P^{\mu}+ p^{\mu})+f^{\chi_{c0}}_m (P^{\mu}- p^{\mu})]
\,.
\end{eqnarray}

The form factors in the Section \ref{II} can be obtained according to the above form factors. For the  $h_c$ and $\chi_{c1}$, we have
\begin{eqnarray}
V^H&=&\frac{m_{B_c}+m_H}{2}g^H\,,\label{trans1}\\
A_0^H&=&\frac{1}{2m_H}a^H_0+\frac{q^2}{4m_H}(a^H_- -a^H_+)\nonumber\\&&+\frac{m^2_{B_c}-m^2_H}{4m_H}(a^H_- +a^H_+)\,,\\
A_1^H&=&\frac{1}{ m_{B_c}+m_H}a^H_0\,,\\
A_2^H&=&-\frac{m_{B_c}+m_H}{2}(a^H_- +a^H_+)\,.
\end{eqnarray}
The form factors of $\chi_{c2}$ in the Section \ref{II} can be obtained through the above transformation plus a factor $m_{B_c}$.

The form factors of $\chi_{c0}$ in the Section \ref{II} can be expressed as
\begin{eqnarray}
f^{\chi_{c0}}_0&=&f^{\chi_{c0}}_p+\frac{q^2}{m^2_{B_c}-m^2_H}f^{\chi_{c0}}_m\,,\\
f^{\chi_{c0}}_+&=&f^{\chi_{c0}}_p\,.\label{transn}
\end{eqnarray}

Thus, the relativistic corrections of the form factors in the Section \ref{II}  will be obtained by the two steps. The first step is to calculate the relativistic corrections from the new defined form factors which do not explicitly depend the $B_c$ meson and $P$-wave charmonium masses. The second step is to calculate the relativistic corrections from the meson  masses in the transformation formulae in Eqs. (\ref{trans1}-\ref{transn}).

\subsection{The direct relativistic corrections to the form factors of $B_c$  into spin-singlet $h_c$ and spin-triplet $\chi_{cJ}$ \label{rc1a}}

In this subsection, we will study the relativistic corrections contributions from the new defined form factors which do not explicitly depend the  meson masses.
Performing the  Taylor expansion of the amplitudes in powers of $k^\mu$ and extracting the quadratic terms in
the series, one then obtain the relativistic corrections at the ${\cal O}(|\bold{k}|^2)$ level.

Normally, one can obtain the short-distance coefficients of the NRQCD relativistic operator matrix elements after performing the relativistic corrections.  To estimate the magnitude of the matrix elements of the relativistic corrections operators, the following estimation formula can be adopted
\begin{eqnarray}
\langle 0|\chi^\dagger_b \left(-\frac{i}{2}  \overleftrightarrow {\bold D}\right)^2  \psi_c
| B_c\rangle&\simeq& |\bold{k}|^2\langle 0|\chi^\dagger_b \psi_c
|B_c\rangle,\label{rcLDMEs}
\end{eqnarray}
where $|\bold{k}|^2$ can be also expressed by the heavy quark relative velocity and heavy quark mass.

Thus the contributions to the form factors of $B_c$ transitions into spin-singlet $h_c$ from the $B_c$ relativistic correction operators  become
\begin{widetext}

\bqa
A_0^{h_c}(q^2)|_{RC1}&=&A_0^{h_c}(q^2)|_{LO}\frac{|\bold{k}|^2 }{m_b^2}[\frac{-15 z^7+519 z^6+1215 z^5+1573 z^4+1263 z^3+689 z^2+129 z+3}{24 z^2 \left((z-1)^2-y^2\right) \left(-y^2 (3 z+2)+5 z^3+8 z^2+9 z+2\right)}\nonumber\\&&+\frac{y^2 \left(12 z^6-211 z^5-137 z^4+74 z^3+186 z^2+105 z+3\right)}{12 (z-1) z^2 \left((z-1)^2-y^2\right) \left(-y^2 (3 z+2)+5 z^3+8 z^2+9 z+2\right)}\nonumber\\&&-\frac{y^4 \left(9 z^4-44 z^3+52 z^2+84 z+3\right)}{24 (z-1) z^2 \left((z-1)^2-y^2\right) \left(-y^2 (3 z+2)+5 z^3+8 z^2+9 z+2\right)}\; ,
 \eqa
 \bqa V^{h_c}(q^2)|_{RC1}&=& V^{h_c}(q^2)|_{LO}\frac{|\bold{k}|^2 }{m_b^2}\frac{y^2 \left(-3 z^2+38 z-3\right)+3 z^4-60 z^3+82 z^2-28 z+3}{24 z^2 (y-z+1) (y+z-1)} ,
  \eqa
\bqa A_1^{h_c}(q^2)|_{RC1}&=& A_1^{h_c}(q^2)|_{LO}\frac{|\bold{k}|^2 }{m_b^2}[\frac{-15 z^6+462 z^5+59 z^4+372 z^3+119 z^2+30 z-3}{24 z^2 \left((z-1)^2-y^2\right) \left(-y^2+5 z^2+2 z+1\right)}\nonumber\\&&-\frac{y^2 \left(y^2 \left(3 z^2-38 z+3\right)-18 z^4+284 z^3+24 z^2+68 z-6\right)}{24 z^2 \left((z-1)^2-y^2\right) \left(-y^2+5 z^2+2 z+1\right)}] ,~~
 \eqa

\bqa A_2^{h_c}(q^2)|_{RC1}&=& A_2^{h_c}(q^2)|_{LO}\frac{|\bold{k}|^2 }{m_b^2}[\frac{-45 z^7+1553 z^6+2747 z^5+3249 z^4+2305 z^3+955 z^2-15 z+3}{24 z^2 \left((z-1)^2-y^2\right) \left(y^2 (1-3 z)+15 z^3+17 z^2+17 z-1\right)}\nonumber\\&&-\frac{y^2 \left(y^2 \left(9 z^3+43 z^2+79 z-3\right)-54 z^5+574 z^4+916 z^3+572 z^2-94 z+6\right)}{24 z^2 \left((z-1)^2-y^2\right) \left(y^2 (1-3 z)+15 z^3+17
   z^2+17 z-1\right)}] .
 \eqa
\end{widetext}

Next, we will study the relativistic corrections to the form factors of $B_c$ transitions into spin-triplet $\chi_{cJ}$.
When one expands the amplitudes in powers of $k^\mu$, the relativistic corrections to the form factors at the ${\cal O}(|\bold{k}|^2)$ level from the $B_c$ meson can be obtained.

We can get the estimations of the relativistic corrections of the form factors according to the above formula in Eq.~\ref{rcLDMEs}, which can be also related to the LO form factors.

Using the estimation formula in Eq.~\ref{rcLDMEs}, the results of relativistic corrections from the $B_c$ meson  to the form factors of $B_c$ transitions into spin-triplet $\chi_{cJ}$ can be written as

\begin{widetext}
\bqa f_+^{\chi_{c0}}(q^{2})|_{RC1}&=&f_+^{\chi_{c0}}(q^{2})|_{LO}\frac{|\bold{k}|^2 }{m_b^2}[\frac{-27 z^8+918 z^7+1338 z^6+574 z^5-220 z^4-926 z^3-1066 z^2-566 z-25}{24 z^2 \left((z-1)^2-y^2\right) \left(-y^4+2 y^2 \left(-2 z^2+z+5\right)+9 z^4+6 z^3-6
   z-9\right)}\nonumber\\&&+\frac{y^6 \left(-\left(3 z^2+2 z+11\right)\right)-3 y^4 \left(3 z^4+38 z^3+156 z^2+154 z+1\right)}{24 z^2 \left((z-1)^2-y^2\right) \left(-y^4+2 y^2 \left(-2
   z^2+z+5\right)+9 z^4+6 z^3-6 z-9\right)}\nonumber\\&&+\frac{y^2 \left(39 z^6-490 z^5-231 z^4+708 z^3+1209 z^2+1030 z+39\right)}{24 z^2 \left((z-1)^2-y^2\right) \left(-y^4+2 y^2 \left(-2 z^2+z+5\right)+9 z^4+6 z^3-6
   z-9\right)}] , \eqa
\bqa f_0^{\chi_{c0}}(q^{2})|_{RC1}&=&f_0^{\chi_{c0}}(q^{2})|_{LO}\frac{|\bold{k}|^2 }{m_b^2}[\frac{-81 z^7+2727 z^6+4851 z^5+5787 z^4+4765 z^3+2789 z^2+641 z+25}{72 z^2 \left((z-1)^2-y^2\right) \left(-y^2 (5 z+3)+9 z^3+9 z^2+11 z+3\right)}\nonumber\\&&-\frac{y^2 \left(y^2 \left(45 z^3-267 z^2-265 z-25\right)-126 z^5+2250 z^4+3604 z^3+2660 z^2+906 z+50\right)}{72 z^2 \left((z-1)^2-y^2\right) \left(-y^2 (5 z+3)+9
   z^3+9 z^2+11 z+3\right)}] ,~~
 \eqa

 \begin{eqnarray}
V^{\chi _{c1}}(q^{2})|_{RC1}
& =& V^{\chi _{c1}}(q^{2})|_{LO}\frac{|
\textbf{k}|^{2} }{m_{b}^{2}}[\frac{-33 z^7+1151 z^6+2411 z^5+3019 z^4+2493 z^3+1437 z^2+313 z-39}{24 z^2 (y-z+1) (y+z-1) \left(y^2 (7 z+5)-11 z^3-15 z^2-17 z-5\right)}\nonumber\\
&&\qquad {} -\tfrac{y^2 \left(3 y^2 \left(7 z^3-33 z^2-19 z+13\right)-54 z^5+938 z^4+1572 z^3+1220 z^2+370 z-78\right)}{24 z^2 (y-z+1) (y+z-1) \left(y^2 (7 z+5)-11 z^3-15 z^2-17
   z-5\right)}] ,
\end{eqnarray}
\bqa A_0^{\chi_{c1}}(q^2)|_{RC1}&=&A_0^{\chi_{c1}}(q^2)|_{LO}\frac{|\bold{k}|^2 }{m_b^2}\frac{y^2 \left(3 z^3-50 z^2+43 z+8\right)-3 z^5+92 z^4-174 z^3+28 z^2-63 z-8}{24 (z-1) z^2 \left((z-1)^2-y^2\right)}\; ,
 \eqa
\bqa A_1^{\chi_{c1}}(q^2)|_{RC1}&=& A_1^{\chi_{c1}}(q^2)|_{LO}\frac{|\bold{k}|^2 }{m_b^2}[\frac{-75 z^7+2003 z^6+4029 z^5+4771 z^4+4151 z^3+2353 z^2+215 z-39}{24 z^2 \left((z-1)^2-y^2\right) \left(-y^2 (9 z+5)+25 z^3+35 z^2+31 z+5\right)}\nonumber\\&&-\frac{y^2 \left(y^2 \left(27 z^3-95 z^2-35 z+39\right)+2 \left(-51 z^5+617 z^4+1062 z^3+750 z^2+125 z-39\right)\right)}{24 z^2 \left((z-1)^2-y^2\right) \left(-y^2
   (9 z+5)+25 z^3+35 z^2+31 z+5\right)}] ,~~
 \eqa

\bqa A_2^{\chi_{c1}}(q^2)|_{RC1}&=& A_2^{\chi_{c1}}(q^2)|_{LO}\frac{|\bold{k}|^2 }{m_b^2}[\frac{-33 z^7+949 z^6-97 z^5+109 z^4+53 z^3-713 z^2-307 z+39}{24 z^2 \left((z-1)^2-y^2\right) \left(y^2 (z+5)+11 z^3-3 z^2-3 z-5\right)}\nonumber\\&&+\frac{y^2 \left(3 y^2 \left(z^3-37 z^2-41 z+13\right)+30 z^5+98 z^4+884 z^3+940 z^2+430 z-78\right)}{24 z^2 \left((z-1)^2-y^2\right) \left(y^2 (z+5)+11 z^3-3 z^2-3
   z-5\right)}] .
 \eqa
\begin{eqnarray}
V^{\chi _{c2}}(q^{2})|_{RC1}
&=& V^{\chi _{c2}}(q^{2})|_{LO}\frac{y^2 \left(-9 z^2+90 z-17\right)+9 z^4-324 z^3-34 z^2-148 z+17}{72 z^2 (y-z+1) (y+z-1)} ,
\end{eqnarray}
\bqa A_0^{\chi_{c2}}(q^2)|_{RC1}&=&A_0^{\chi_{c2}}(q^2)|_{LO}\frac{|\bold{k}|^2 }{m_b^2}\frac{- \left(y^2 \left(-9 z^2+108 z+13\right)+9 z^4-342 z^3-124 z^2-202 z-13\right)}{72 z^2 \left((z-1)^2-y^2\right)}\; ,
 \eqa
\bqa A_1^{\chi_{c2}}(q^2)|_{RC1}&=& A_1^{\chi_{c2}}(q^2)|_{LO}\frac{|\bold{k}|^2 }{m_b^2}[\frac{-27 z^7+945 z^6+2433 z^5+3229 z^4+2655 z^3+1411 z^2+123 z-17}{24 z^2 \left((z-1)^2-y^2\right) \left(-y^2 (5 z+3)+9 z^3+17 z^2+19 z+3\right)}\nonumber\\&&+\frac{-27 z^7+945 z^6+2433 z^5+3229 z^4+2655 z^3+1411 z^2+123 z-17}{24 z^2 \left((z-1)^2-y^2\right) \left(-y^2 (5 z+3)+9 z^3+17 z^2+19 z+3\right)}] ,~~
 \eqa

\bqa A_2^{\chi_{c2}}(q^2)|_{RC1}&=& A_2^{\chi_{c2}}(q^2)|_{LO}\frac{|\bold{k}|^2 }{m_b^2}\frac{y^2 \left(3 z^3-85 z^2-95 z+17\right)-3 z^5+115 z^4+330 z^3+238 z^2+105 z-17}{24
   z^2 (z+3) \left((z-1)^2-y^2\right)} .
 \eqa
 \end{widetext}

Next, we will study the relativistic corrections contributions from the charm quark relative velocity inside the $P$-wave charmonium.
Performing the  Taylor expansion of the amplitudes in powers of $k'^\mu$ and extracting the quadratic terms in
the series, one then obtain the relativistic corrections at the ${\cal O}(|\bold{k}'|^2)$ level.

The magnitudes of the matrix elements of the  $P$-wave charmonium relativistic corrections operators are estimated as
\bqa
&&\langle h_c(\varepsilon^*)|
\psi^{\dagger}(-\frac{i}{2}{\overleftrightarrow{ { D}^i}})\left(-\frac{i}{2}  \overleftrightarrow {\bold D}\right)^2\chi|0\rangle
\nonumber\\&&\simeq |\bold{k}'|^2\langle h_c(\varepsilon^*)|
\psi^{\dagger}(-\frac{i}{2}{\overleftrightarrow{ { D}^i}})\chi|0\rangle.
\\
&&\langle \chi_{c0}|
\psi^{\dagger}(-\frac{i}{2}{\overleftrightarrow{ {\bold D}}}\cdot \bfsigma)\left(-\frac{i}{2}  \overleftrightarrow {\bold D}\right)^2\chi|0\rangle
\nonumber\\&&\simeq |\bold{k}'|^2\langle \chi_{c0}|
\psi^{\dagger}(-\frac{i}{2}{\overleftrightarrow{ {\bold D}}}\cdot \bfsigma)\chi|0\rangle.
\eqa
\bqa
&&\langle \chi_{c1}(\varepsilon^*)|
\psi^{\dagger}(-\frac{i}{2}{\overleftrightarrow{ {\bold D}}}\times \bfsigma)^i\left(-\frac{i}{2}  \overleftrightarrow {\bold D}\right)^2\chi|0\rangle
\nonumber\\&&\simeq |\bold{k}'|^2\langle \chi_{c1}(\varepsilon^*)|
\psi^{\dagger}(-\frac{i}{2}{\overleftrightarrow{ {\bold D}}}\times \bfsigma)^i\chi|0\rangle.
\eqa
\bqa
&&\langle \chi_{c2}(\varepsilon^*)|
\psi^{\dagger}(-\frac{i}{2}{\overleftrightarrow{ { D}^i}} \sigma^j)\left(-\frac{i}{2}  \overleftrightarrow {\bold D}\right)^2\chi|0\rangle
\nonumber\\&&\simeq |\bold{k}'|^2\langle \chi_{c2}(\varepsilon^*)|
\psi^{\dagger}(-\frac{i}{2}{\overleftrightarrow{ { D}^i}} \sigma^j)\chi|0\rangle.
 \eqa

Thus the contributions to the form factors of $B_c$ transitions into spin-singlet $h_c$  from the $P$-wave charmonium relativistic corrections operators  become
\begin{widetext}
 \bqa V^{h_c}(q^2)|_{RC'1}&=& V^{h_c}(q^2)|_{LO}\frac{|\bold{k}'|^2 }{m_b^2}\frac{2 \left(y^2-9 z^2+14 z-1\right)}{3 z^2 \left((z-1)^2-y^2\right)},
 \eqa
\bqa A_0^{h_c}(q^2)|_{RC'1}&=&A_0^{h_c}(q^2)|_{LO}\frac{|\bold{k}'|^2 }{m_b^2}[\frac{-161 z^5+166 z^4+540 z^3+818 z^2+229 z+8}{6 z^2 \left((z-1)^2-y^2\right) \left(-y^2 (3 z+2)+5 z^3+8 z^2+9 z+2\right)}\nonumber\\&&+\frac{2 y^2 \left(49 z^4-107 z^3+59 z^2+111 z+8\right)-y^4 \left(z^2+z+8\right)}{6 (z-1) z^2 \left((z-1)^2-y^2\right) \left(-y^2 (3 z+2)+5 z^3+8 z^2+9 z+2\right)}]\; ,
 \eqa
\bqa A_1^{h_c}(q^2)|_{RC'1}&=& A_1^{h_c}(q^2)|_{LO}\frac{|\bold{k}'|^2 }{m_b^2}\frac{-2 y^4+4 y^2 \left(2 z^2-6 z+1\right)-70 z^4+64 z^3+48 z^2+24 z-2}{3 z^2 \left((z-1)^2-y^2\right) \left(-y^2+5 z^2+2 z+1\right)} ,~~
 \eqa

\bqa A_2^{h_c}(q^2)|_{RC'1}&=& A_2^{h_c}(q^2)|_{LO}\frac{|\bold{k}'|^2 }{m_b^2}[\frac{2 \left(y^4 (1-3 z)-119 z^5+161 z^4+326 z^3+430 z^2+z+1\right)}{3 z^2 \left((z-1)^2-y^2\right) \left(y^2 (1-3 z)+15 z^3+17 z^2+17 z-1\right)}\nonumber\\&&-\frac{4 y^2 \left(3 z^3+57 z^2-z+1\right)}{3 z^2 \left((z-1)^2-y^2\right) \left(y^2 (1-3 z)+15 z^3+17 z^2+17 z-1\right)}] .
 \eqa
\end{widetext}

The contributions to the form factors of $B_c$ transitions into spin-triplet $\chi_{cJ}$  from the $P$-wave charmonium relativistic corrections operators  become
\begin{widetext}
\bqa f_+^{\chi_{c0}}(q^{2})|_{RC'1}&=&f_+^{\chi_{c0}}(q^{2})|_{LO}\frac{|\bold{k}'|^2 }{m_b^2}[\frac{-5 y^6-189 z^6+532 z^5+719 z^4+176 z^3-479 z^2-772 z+13}{6 z^2 \left((z-1)^2-y^2\right) \left(-y^4+2 y^2 \left(-2 z^2+z+5\right)+9 z^4+6 z^3-6 z-9\right)}\nonumber\\&&+\frac{y^4 \left(25 z^2-80 z+23\right)+y^2 \left(41 z^4-460 z^3-82 z^2+852 z-31\right)}{6 z^2 \left((z-1)^2-y^2\right) \left(-y^4+2 y^2 \left(-2 z^2+z+5\right)+9
   z^4+6 z^3-6 z-9\right)}] , \eqa
\bqa f_0^{\chi_{c0}}(q^{2})|_{RC'1}&=&f_0^{\chi_{c0}}(q^{2})|_{LO}\frac{|\bold{k}'|^2 }{m_b^2}[\frac{-567 z^6+840 z^5+3529 z^4+4776 z^3+3515 z^2+720 z-13}{18 z^2 (z+1) \left((z-1)^2-y^2\right) \left(-y^2 (5 z+3)+9 z^3+9 z^2+11 z+3\right)}\nonumber\\&&-\frac{y^2 \left(y^2 \left(3 z^2+48 z+13\right)-122 z^4+1104 z^3+2084 z^2+672 z-26\right)}{18 z^2 (z+1) \left((z-1)^2-y^2\right) \left(-y^2 (5 z+3)+9 z^3+9 z^2+11
   z+3\right)}] ,~~
 \eqa

 \bqa V^{\chi_{c1}}(q^2)|_{RC'1}&=& V^{\chi_{c1}}(q^2)|_{LO}\frac{|\bold{k}'|^2 }{m_b^2}[\frac{y^4 (9 z+3)-227 z^5+447 z^4+1058 z^3+1422 z^2+497 z+3}{6 z^2 (y-z+1) (y+z-1) \left(y^2 (7 z+5)-11 z^3-15 z^2-17 z-5\right)}\nonumber\\&&+\frac{y^2 \left(45 z^3-261 z^2-253 z-3\right)}{3 z^2 (y-z+1) (y+z-1) \left(y^2 (7 z+5)-11 z^3-15 z^2-17 z-5\right)}],
 \eqa
\bqa A_0^{\chi_{c1}}(q^2)|_{RC'1}&=&A_0^{\chi_{c1}}(q^2)|_{LO}\frac{|\bold{k}'|^2 }{m_b^2}\frac{y^2 \left(6 z^2-15 z-1\right)-74 z^4+75 z^3+33 z^2-35 z+1}{6 (z-1) z^2 (z+1) \left((z-1)^2-y^2\right)}\; ,
 \eqa
\bqa A_1^{\chi_{c1}}(q^2)|_{RC'1}&=& A_1^{\chi_{c1}}(q^2)|_{LO}\frac{|\bold{k}'|^2 }{m_b^2}[\frac{-y^4 (z-3)-449 z^5+1183 z^4+2678 z^3+2494 z^2+491 z+3}{6 z^2 \left((z-1)^2-y^2\right) \left(-y^2 (9 z+5)+25 z^3+35 z^2+31 z+5\right)}\nonumber\\&&-\frac{y^2 \left(15 z^3+385 z^2+245 z+3\right)}{3 z^2 \left((z-1)^2-y^2\right) \left(-y^2 (9 z+5)+25 z^3+35 z^2+31 z+5\right)}] ,~~
 \eqa

\bqa A_2^{\chi_{c1}}(q^2)|_{RC'1}&=& A_2^{\chi_{c1}}(q^2)|_{LO}\frac{|\bold{k}'|^2 }{m_b^2}[-\frac{y^4 (39 z+3)+347 z^5-1057 z^4-34 z^3+254 z^2+487 z+3}{6 z^2 \left((z-1)^2-y^2\right) \left(y^2 (z+5)+11 z^3-3 z^2-3 z-5\right)}\nonumber\\&&-\frac{y^2 \left(63 z^3+75 z^2-263 z-3\right)}{3 z^2 \left((z-1)^2-y^2\right) \left(y^2 (z+5)+11 z^3-3 z^2-3 z-5\right)}] .
 \eqa
\bqa V^{\chi_{c2}}(q^2)|_{RC'1}&=& V^{\chi_{c2}}(q^2)|_{LO}\frac{|\bold{k}'|^2 }{m_b^2}\frac{4 \left(y^2 (3 z+2)+13 z^3-21 z^2-40 z-2\right)}{9 z^2 (z+1) (y-z+1) (y+z-1)},
 \eqa
\bqa A_0^{\chi_{c2}}(q^2)|_{RC'1}&=&A_0^{\chi_{c2}}(q^2)|_{LO}\frac{|\bold{k}'|^2 }{m_b^2}\frac{- \left(y^2 \left(15 z^2+21 z+16\right)+49 z^4-27 z^3-253 z^2-153 z-16\right)}{9z^2 (z+1)^2 \left((z-1)^2-y^2\right)}\; ,
 \eqa
\bqa A_1^{\chi_{c2}}(q^2)|_{RC'1}&=& A_1^{\chi_{c2}}(q^2)|_{LO}\frac{|\bold{k}'|^2 }{m_b^2}[\frac{4 \left(y^4 (3 z+2)+2 \left(-8 z^5+16 z^4+67 z^3+99 z^2+25 z+1\right)\right)}{3 z^2 \left((z-1)^2-y^2\right) \left(-y^2 (5 z+3)+9 z^3+17 z^2+19 z+3\right)}\nonumber\\&&-\frac{4 y^2 \left(3 z^3+64 z^2+53 z+4\right)}{3 z^2 \left((z-1)^2-y^2\right) \left(-y^2 (5 z+3)+9 z^3+17 z^2+19 z+3\right)}] ,~~
 \eqa

\bqa A_2^{\chi_{c2}}(q^2)|_{RC'1}&=& A_2^{\chi_{c2}}(q^2)|_{LO}\frac{|\bold{k}'|^2 }{m_b^2}\frac{-4 \left(y^2 (z+2)+11 z^3+9 z^2-30 z-2\right)}{3 z^2 (z+3) \left((z-1)^2-y^2\right)}.
 \eqa
 \end{widetext}

 \subsection{The corrections from the masses of $B_c$ meson  and $P$-wave charmonium\label{bcmas}}

Through the transformation formulae in Eqs. (\ref{trans1}-\ref{transn}), one can easily get
the relativistic corrections from the $B_c$ meson and $P$-wave charmonium masses. The relativistic corrections of the  $h_c$ and $\chi_{c1}$ form factors from the
$B_c$ meson  mass are

\begin{eqnarray}
V^H(q^2)|_{RC2}&=&\frac{z+1}{2 z (3 z+1)}\frac{|\bold{k}|^2 }{m_b^2} V^H(q^2)|_{LO}\,,\\
A_0^H(q^2)|_{RC2}&=&-\frac{ (z+1)^2}{4 z^2 (3 z+1)}\frac{|\bold{k}|^2 }{m_b^2}A_2^H(q^2)|_{LO}\,,\\
A_1^H(q^2)|_{RC2}&=&-\frac{z+1}{2 z (3 z+1)}\frac{|\bold{k}|^2 }{m_b^2}A_1^H(q^2)|_{LO}\,,\\
A_2^H(q^2)|_{RC2}&=&\frac{z+1}{2 z (3
   z+1)}\frac{|\bold{k}|^2 }{m_b^2}A_2^H(q^2)|_{LO}\,,
\end{eqnarray}
here $H$ only denotes $h_c$ or $\chi_{c1}$.

 The relativistic corrections of the $\chi_{c2}$ form factors from the
$B_c$ meson  mass are
\begin{eqnarray}
V^{\chi_{c2}}(q^2)|_{RC2}&=&\frac{-1}{(z+1) (3 z+1)} \frac{|\bold{k}|^2 }{m_b^2}V^{\chi_{c2}}|_{LO}\,,\\
A_0^{\chi_{c2}}(q^2)|_{RC2}&=&\frac{ -\left(5 z^2+2 z+1\right)}{8 z^2 (z+1) (3 z+1)}\frac{|\bold{k}|^2 }{m_b^2}A_2^{\chi_{c2}}|_{LO},\\
A_1^{\chi_{c2}}(q^2)|_{RC2}&=&\frac{-(2 z+1)}{z (z+1) (3 z+1)}\frac{|\bold{k}|^2 }{m_b^2}A_1^{\chi_{c2}}|_{LO}\,,\\
A_2^{\chi_{c2}}(q^2)|_{RC2}&=&\frac{-1}{(z+1) (3 z+1)}\frac{|\bold{k}|^2 }{m_b^2}A_2^{\chi_{c2}}|_{LO}\,.
\end{eqnarray}

 The relativistic corrections of the $\chi_{c0}$ form factors from the
$B_c$ meson  mass are
\begin{eqnarray}
&&f^{\chi_{c0}}_0(q^2)|_{RC2}=\frac{(z+1)^2}{(z-1) z (3 z+1)}\frac{|\bold{k}|^2 }{m_b^2}
\nonumber\\&&\quad\quad\quad\quad\quad\quad\times(f^{\chi_{c0}}_0(q^2)|_{LO}-f^{\chi_{c0}}_+(q^2)|_{LO})\,,\quad\\
&&f^{\chi_{c0}}_+(q^2)|_{RC2}=0\,.
\end{eqnarray}

The relativistic corrections of the $P$-wave charmonium form factors from the
$P$-wave charmonium  mass are
\begin{eqnarray}
V^H(q^2)|_{RC'2}&=&\frac{1}{z (3 z+1)}\frac{|\bold{k}'|^2 }{m_b^2} V^H(q^2)|_{LO}\,,\\
A_1^H(q^2)|_{RC'2}&=&-\frac{1}{z (3 z+1)}\frac{|\bold{k}'|^2 }{m_b^2}A_1^H(q^2)|_{LO}\,,\\
A_2^H(q^2)|_{RC'2}&=&\frac{1}{z (3 z+1)}\frac{|\bold{k}'|^2 }{m_b^2}A_2^H(q^2)|_{LO}\,,
\end{eqnarray}
here $H$ only denotes $h_c$, $\chi_{c1}$ or $\chi_{c2}$, and
\begin{eqnarray}
&&A_0^H(q^2)|_{RC'2}=\frac{|\bold{k}'|^2 }{m_b^2} [\frac{ \left(5 z^2+2 z+1\right)}{2 (z-1) z^2 (3 z+1)}A_0^H(q^2)|_{LO}\nonumber\\
&&\quad+\frac{(3z+1)A_1^H(q^2)|_{LO} +(z-1)A_2^H(q^2)|_{LO}}{-3 z^3+2 z^2+z}]\,.
\end{eqnarray}
For $\chi_{c0}$, we have
\begin{eqnarray}
&&f^{\chi_{c0}}_0(q^2)|_{RC'2}=-\frac{4 }{(z-1) (3 z+1)}\frac{|\bold{k}'|^2 }{m_b^2}
\nonumber\\&&\quad\quad\quad\quad\quad\quad\times(f^{\chi_{c0}}_0(q^2)|_{LO}-f^{\chi_{c0}}_+(q^2)|_{LO})\,,\quad\\
&&f^{\chi_{c0}}_+(q^2)|_{RC'2}=0\,.
\end{eqnarray}

\begin{table*}[ht]
\begin{center}
\caption{The form factors of the $B_c$ meson into $P$-wave charmonium  at
$q^2=0$ in different approaches. In NRQCD approach, the scale is set at the heavy quark mass $m_b$,
and we adopt  $m_b=(4.7\pm0.1)$GeV and $m_c=(1.5\pm0.1)$GeV. The first column uncertainty is from the choice of $m_b$, while the second column uncertainty is from the choice of $m_c$.
}
\begin{tabular}{ccccc}
 \hline\hline
Form factors& NRQCD LO& NRQCD LO+RC & LFQM~\cite{Wang:2009mi} & QCD SR~\cite{Azizi:2009ny}
   \\
 \hline
$V^{h_{c}}(0)$
& $0.24^{+0.03+0.01}_{-0.02-0.00}$ & $0.24^{+0.02+0.00}_{-0.02-0.00}$ &$0.24$ & $0.48$ \\
$A_{0}^{h_{c}}(0)$
& $1.63^{+0.19+0.10}_{-0.17-0.08}$ & $2.11^{+0.28+0.15}_{-0.23-0.10}$ &$0.64$ & $0.03$ \\
$A_{1}^{h_{c}}(0)$
& $0.13^{+0.02+0.00}_{-0.01-0.00}$ & $0.15^{+0.02+0.01}_{-0.01-0.00}$ &$0.14$ & $0.08$ \\
$A_{2}^{h_{c}}(0)$
& $-2.66^{+0.43+0.39}_{-0.35-0.50}$ & $-3.51^{+0.60+0.53}_{-0.50-0.67}$ &$-1.14$ & $0.21$ \\
$f_{0}^{\chi _{c0}}(0)=f_{+}^{\chi _{c0}}(0)$
& $1.25^{+0.14+0.06}_{-0.12-0.03}$ & $1.65^{+0.21+0.08}_{-0.17-0.05}$ &$0.47$ & $0.67$ \\
$V^{\chi _{c1}}(0)$
& $3.49^{+0.49+0.45}_{-0.40-0.35}$ & $4.40^{+0.68+0.62}_{-0.56-0.47}$ &$1.28$ & $0.47$ \\
$A_{0}^{\chi _{c1}}(0)$
& $0.12^{+0.01+0.00}_{-0.01-0.01}$ & $0.17^{+0.01+0.01}_{-0.02-0.01}$ &$0.13$ & $0.03$ \\
$A_{1}^{\chi _{c1}}(0)$
& $0.66^{+0.06+0.00}_{-0.06-0.00}$ & $0.81^{+0.01+0.01}_{-0.07-0.00}$ &$0.24$ & $0.08$ \\
$A_{2}^{\chi _{c1}}(0)$
& $1.67^{+0.21+0.13}_{-0.18-0.10}$ & $2.03^{+0.27+0.17}_{-0.23-0.13}$ &$0.53$ & $0.21$ \\
$V^{\chi _{c2}}(0)$
& $5.89^{+1.00+1.29}_{-0.83-1.02}$ & $6.74^{+1.20+1.53}_{-1.02-1.20}$ &$1.34$ & \\
$A_{0}^{\chi _{c2}}(0)$
& $1.80^{+0.27+0.29}_{-0.23-0.23}$ & $2.39^{+0.40+0.40}_{-0.32-0.31}$ &$0.86$ & \\
$A_{1}^{\chi _{c2}}(0)$
& $1.95^{+0.30+0.31}_{-0.25-0.25}$ & $2.46^{+0.43+0.42}_{-0.33-0.33}$ &$0.81$ & \\
$A_{2}^{\chi _{c2}}(0)$
& $2.24^{+0.34+0.38}_{-0.29-0.30}$ & $2.59^{+0.42+0.47}_{-0.35-0.37}$ &$0.68
$ \\
\hline
\end{tabular}\label{table1}
\end{center}
\end{table*}

\subsection{The heavy bottom quark limit\label{limit}}

Since $z=
m_c/m_b\approx 0.3$  is small, it is convenient to study the form factor in the heavy bottom quark limit. In the heavy bottom quark limit $m_b \to \infty$ and $z=
m_c/m_b\rightarrow 0$, one can obtain the relations among form factors.
In this limit, the form factors at the maximum recoil point with $q^2=0$ become
 \bqa
V^{h_c}(0)|_{LO}^{m_b \rightarrow\infty}&=&\frac{8 \sqrt{2} \pi   C_F \alpha _s\psi(0)_{B_c}\psi'(0)_{h_c} }{z^{5/2}m_b^4 } ,
\\ V^{\chi_{c1}}(0)|_{LO}^{m_b \rightarrow\infty}&=&\frac{40 \pi  C_F \alpha _s\psi(0)_{B_c}\psi'(0)_{\chi_{c1}}}{z^{5/2} m_b^4} ,
\\ V^{\chi_{c2}}(0)|_{LO}^{m_b \rightarrow\infty}&=&\frac{96 \sqrt{2} \pi   C_F \alpha _s \psi(0)_{B_c}\psi'(0)_{\chi_{c2}}}{z^{3/2}m_b^4} ,\nonumber\\
 \eqa
 \bqa
 A_2^{H}(0)|_{LO}^{m_b \rightarrow\infty}&=&A_1^{H}(0)|_{LO}^{m_b \rightarrow\infty}
=V^{H}(0)|_{LO}^{m_b \rightarrow\infty} ,
 \eqa
 where $H$ denotes one of the charmonia $h_c$, $\chi_{c1}$ and $\chi_{c2}$. From the above formulae,
 the form factors are not independent in the heavy quark limit  and their relations become simple.

The form factors of  $\chi_{c0}$ at the maximum recoil point  in the heavy bottom quark limit become
\bqa
f_+^{\chi_{c0}}(0)|_{LO}^{m_b \rightarrow\infty}&=&\frac{24 \sqrt{6} \pi  C_F \alpha _s \psi(0)_{B_c}\psi'(0)_{\chi_{c0}} }{z^{5/2}m_b^4 },\nonumber\\\\
f_0^{\chi_{c0}}(0)&=&f_+^{\chi_{c0}}(0)\;\label{ff1} ,
 \eqa
where the last formulae is valid for all orders.

The relativistic corrections bring about the similar relations,
which are in consistent with the predictions of the heavy quark effect theory~\cite{Stech:1995ec} and the large energy effective theory~\cite{Charles:1998dr}. In the heavy quark limit, the form factors can be expressed by several simple parameters. Some form factors become identical. The form factors considering the relativistic corrections become

\bqa
V^{h_c}(0)|_{RC1}^{m_b \rightarrow\infty}&=&-\frac{1}{8}\frac{|\bold{k}|^2 }{z^2m_b^2}V^{h_c}(0)|_{LO}^{m_b \rightarrow\infty},
\\ V^{\chi_{c1}}(0)|_{RC1}^{m_b \rightarrow\infty}&=&-\frac{13}{40}\frac{|\bold{k}|^2 }{z^2m_b^2}V^{\chi_{c1}}(0)|_{LO}^{m_b \rightarrow\infty} ,
\\ V^{\chi_{c2}}(0)|_{RC1}^{m_b \rightarrow\infty}&=&-\frac{17}{72}\frac{|\bold{k}|^2 }{z^2m_b^2}V^{\chi_{c2}}(0)|_{LO}^{m_b \rightarrow\infty} ,
\\f_+^{\chi_{c0}}(0)|_{RC1}^{m_b \rightarrow\infty}&=&\frac{25}{216}\frac{|\bold{k}|^2 }{z^2m_b^2}f_+^{\chi_{c0}}(0)|_{LO}^{m_b \rightarrow\infty},
 \eqa
 and
 \bqa
V^{h_c}(0)|_{RC'1}^{m_b \rightarrow\infty}&=&-\frac{2}{3}\frac{|\bold{k}'|^2 }{z^2m_b^2}V^{h_c}(0)|_{LO}^{m_b \rightarrow\infty},
\\ V^{\chi_{c1}}(0)|_{RC'1}^{m_b \rightarrow\infty}&=&\frac{1}{10}\frac{|\bold{k}'|^2 }{z^2m_b^2}V^{\chi_{c1}}(0)|_{LO}^{m_b \rightarrow\infty} ,
\\ V^{\chi_{c2}}(0)|_{RC'1}^{m_b \rightarrow\infty}&=&\frac{8}{9}\frac{|\bold{k}'|^2 }{z^2m_b^2}V^{\chi_{c2}}(0)|_{LO}^{m_b \rightarrow\infty} ,
\\f_+^{\chi_{c0}}(0)|_{RC'1}^{m_b \rightarrow\infty}&=&\frac{104}{3\sqrt{3}}\frac{|\bold{k}'|^2 }{z^2m_b^2}f_+^{\chi_{c0}}(0)|_{LO}^{m_b \rightarrow\infty},
 \eqa
 \bqa
 A_2^{H}(0)|_{RC^{(')}1}^{m_b \rightarrow\infty}=A_1^{H}(0)|_{RC^{(')}1}^{m_b \rightarrow\infty}
=V^{H}(0)|_{RC^{(')}1}^{m_b \rightarrow\infty} ,&&
 \eqa

\section{Phenomenological discussions\label{IV}}

We  obtained the relativistic corrections to the form factors of  $B_c$ into $P$-wave charmonium in the last section. Before employing the form factors, we first summarize and give the values of the form factors at the maximum recoil point. And then we will employ the form factors into the exclusive two-body decays $B_c \to h_c +\pi $, $B_c \to h_c +K $, $B_c \to \chi_{cJ} +\pi $,  and $B_c \to \chi_{cJ} +K $.

The  values  of the related parameters are adopted as follows
~\cite{Olive:2016xmw}
\begin{center}
$m_{B_c}=6.276${GeV},~~$m_{h_c}=3.525${GeV},~\\
$m_{\chi_{c0}}=3.415${GeV},~~$m_{\chi_{c1}}=3.511${GeV},~\\
$m_{\chi_{c2}}=3.556${GeV},~~$G_F=1.16637\times10^{-5}\mathrm{GeV}^{-2}$.
\end{center}
The strong
coupling constant is set at the Z-boson point with  $\alpha_s(m_Z)=0.1185$ where $m_Z=91.1876${GeV}~\cite{Olive:2016xmw}, so one can run the coupling to other scales. In the paper, we will set the scale to the bottom quark mass, and we have $\alpha_s(m_b=4.7\mathrm{GeV})=0.218$ at two-loop evolution with active flavor $n_f=5$.
The Cabibbo-Kobayashi-Maskawa matrix-elements  are set as $|V_{ud}|=0.974$, $|V_{cb}|=(40.5\pm 1.5)\times 10^{-3}$ and $|V_{us}|=0.2248$~\cite{Olive:2016xmw}.
The decay constants of the light mesons are set as $f_\pi=130.4$MeV and $f_K=156.1$MeV.

 The heavy quark mass is adopted as $m_c=1.5\pm0.1${GeV}~\cite{Zhu:2015jha,Jiang:2015jma} and $m_b=4.7\pm0.1${GeV}~\cite{Wang:2015bka}.
This choice of the heavy quark masses is from the nonrelativistic approximation, since the heavy quark relativistic velocity is small in the heavy quarkonium. For the ground state of the $B_c$ family,  we have $m_{B_c}=\sqrt{m_c^2-k^2}+\sqrt{m_b^2-k^2}\simeq m_c+m_b+\frac{\bold k^2}{2m_c}+\frac{\bold k^2}{2m_b}$. For heavy quarkonium, we have $m_{H}=2\sqrt{m_Q^2-{k'}^2}\simeq 2m_Q+\frac{{{\bold k}'}^2}{2m_Q}$.

To evaluate the contributions from the relativistic correction operators, one should
estimate the magnitude of the quark relative momentum in the $B_c$ meson and $P$-wave charmonium. Employing
${\bold k}=m_{red}{\bold v}$,  we have $|{\bold k}|^2=m^2_{red}|{\bold v}|^2=m_b^2m_c^2|{\bold v}|^2/(m_b+m_c)^2$. The reduced quark relative velocity squared in the $B_c$ meson is adopted as
 $|{\bold v}|^2\simeq 0.186$ in Ref.~\cite{Zhu:2017lqu}.  Employing
${\bold k}'=m_c{\bold v}'/2$,  we have $|{\bold k}'|^2=m^2_c|{\bold v}'|^2/4$. The charm quark relative velocity squared in the  $P$-wave charmonium is adopted as
 $|{\bold v}'|^2\simeq 0.2$.

The wave functions at the origin of heavy quarkonium and the $B_c$ meson have been
evaluated in the QCD-motivated Buchmuller-Tye potential~\cite{Buchmuller:1980su,Eichten:1995ch}
\bqa
\psi(0)_{B_c}&=&\frac{1}{\sqrt{4\pi}}R(0)_{B_c}=0.3615\;,\\
\psi'(0)_{h_c}&=&\frac{\sqrt{3}}{\sqrt{4\pi}}R'(0)_{h_c}=0.1338\;,\\
\psi'(0)_{\chi_{cJ}}&=&\frac{\sqrt{3}}{\sqrt{4\pi}}R'(0)_{\chi_{cJ}}=0.1338\;.
 \eqa

\begin{figure}[th]
\begin{center}
\includegraphics[width=0.45\textwidth]{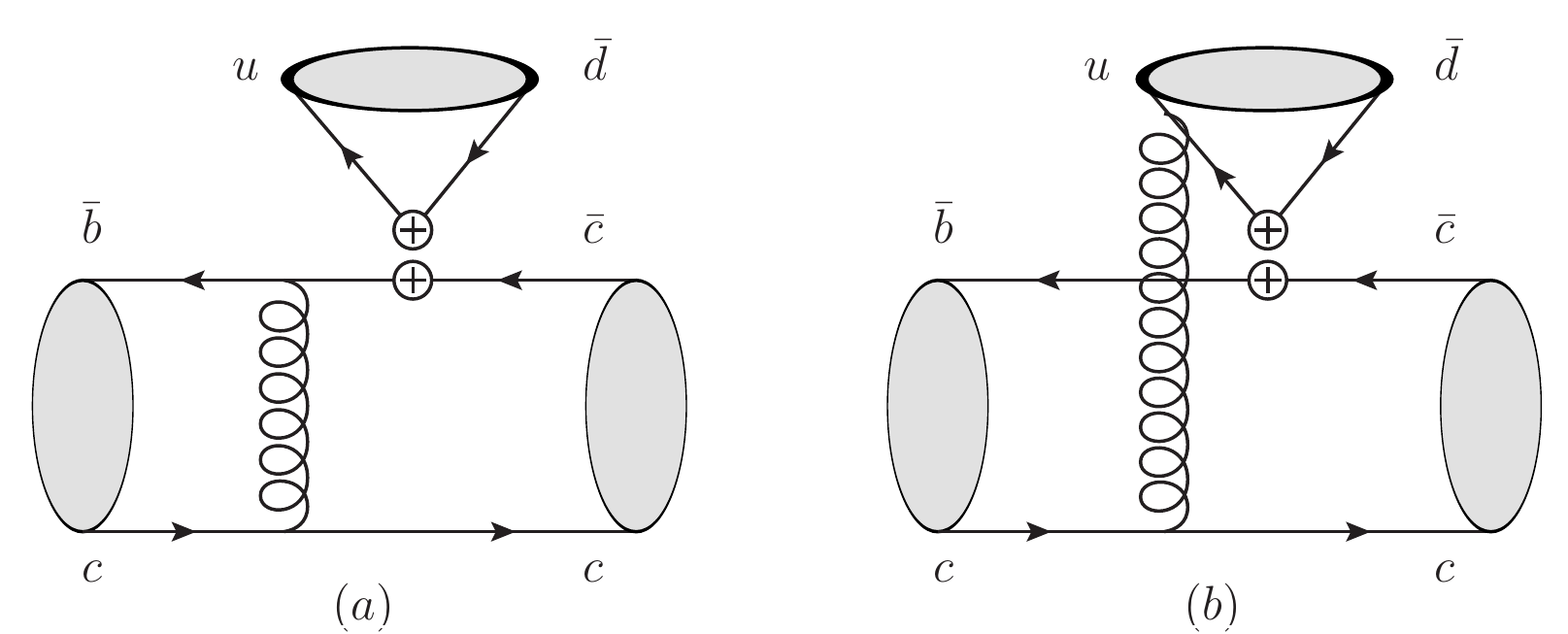}
\end{center}
    \vskip -0.7cm \caption{Typical Feynman diagrams for the $B_c$ meson decays into $P$-wave charmonium and a light meson.}\label{Fig-diagram}
\end{figure}

 Inputting the values of the related parameters in the form factors, we obtained the LO and relativistic correction values of the form factors at the maximum recoil point with $q^2=0$, which have been given in Tab.~\ref{table1}. Comparing to the predictions of the form factors in LFQM and QCD SR, the NRQCD results are consistent with LFQM results plus an additional enhancement factor. 
Extrapolating the form factors to the minimum momentum recoil region where the strong interaction among gluons and quarks becomes nonperturbative,
the pole  model are generally adopted in many literatures~\cite{Kiselev:1999sc,Wang:2008xt,Zhu:2017lqu}. Let us denote  anyone of the form factors as $F^{H}(q^2)$.  Thus the $q^2$ distribution of the form factor can be
parametrized as
\begin{equation}\label{pole mass}
    F^H(q^2)=\frac{F^H(0)}{1-\frac{q^2}{m^2_{\mathrm{pole}}}-
    \beta\frac{ q^4}{m^4_{\mathrm{pole}}}}\; ,
\end{equation}
where  the effective pole mass $m_{\mathrm{pole}}$  is usually set to the value between the heavy quark mass and the initial hadron mass. The free parameter $\beta$  is usually dependent on the certain transition channel.

Through the study of $B^+_c \to K^+ K^- \pi^+$ using a dataset of integrated luminosities of 3$fb^{-1}$  at centre-of-mass energies of 7TeV and 8TeV, the LHCb
Collaboration have observed the evidence of the decay $B_c^+\to \chi_{c0} \pi^+$
for the first time. It is also the first time to observe the $B_c$ decays to $P$-wave charmonium in experiment. The data give $\frac{\sigma(B_c^+)}{\sigma(B^+)}\times {\cal B}(B_c^+\to \chi_{c0} \pi^+)=(9.8_{-3.0}^{+3.4}(stat)\pm0.8(syst))\times 10^{-6}$ with a significance of 4.0 standard deviations.

 In the following, we will employ the form factors into the   exclusive $B_c$ two-body decays into a $P$-wave charmonium and a light meson such as $\pi$ and $K$.
In weak effective theory, the decay amplitudes of $B_c^+\to h_c(\chi_{c1} )+\pi^+$ can be factorized as~\cite{Qiao:2012hp}
 \begin{widetext}
\begin{eqnarray}
{\cal M}(B^+_{c}\to  h_c(\chi_{c1} ) +\pi^+)&=&\sum_i\frac{G_F}{\sqrt{2}}V_{cb}^*V_{ud}C_{i}(\mu)
\langle   h_c(\chi_{c1} ) +\pi^+\vert {\cal O}_{i}\vert B_{c}^{+}\rangle\nonumber\\
&=&\frac{G_F}{\sqrt{2}}V_{cb}^*V_{ud}\left[C_{0}(\mu)\langle   h_c(\chi_{c1} ) +\pi^+\vert {\cal O}_{0}\vert B_{c}^{+}\rangle+C_{8}(\mu)\langle   h_c(\chi_{c1} ) +\pi^+\vert {\cal O}_{8}\vert B_{c}^{+}\rangle\right]
\nonumber\\
&=&\frac{G_F}{\sqrt{2}}V_{cb}^*V_{ud}\left(C_{0}(\mu)\langle h_c(\chi_{c1} )\vert \bar b
\gamma^{\mu}(1-\gamma_{5})c\vert B_{c}^{+}\rangle\langle
\pi^{+}\vert \bar u\gamma_{\mu}(1-\gamma_{5})d\vert 0\rangle\right.
\nonumber\\&&\left.+C_{8}(\mu)\langle   h_c(\chi_{c1} ) +\pi^+\vert {\cal O}_{8}\vert B_{c}^{+}\rangle\right)\,,
\label{pi2}
\end{eqnarray}
 \end{widetext}

where ${\cal O}_{0}$ is
the color-singlet weak effective four-fermion operator, and ${\cal O}_{8}$ is
the color-octet weak effective four-fermion operator. $C_{i}(\mu)$ is  the corresponding Wilson coefficients.  In contrast to the conventional four-fermion operators ${\cal O}_{1,2}(\mu)$,
this operator basis is very convenient in the calculation.
\begin{eqnarray}
{\cal O}_{0}&=&\bar d_{\alpha}\gamma^{\mu}(1-\gamma_{5})u_{\alpha}\bar
c_{\beta}\gamma_{\mu}
(1-\gamma_{5})b_{\beta}\,,\\
{\cal O}_{8}&=&\bar
d_{\alpha}T^{A}_{\alpha\beta}\gamma^{\mu}(1-\gamma_{5})u_{\beta}\bar
c_{\rho}T_{\rho\lambda}^{A}\gamma_{\mu}(1-\gamma_{5})b_{\lambda}\,.\nonumber\\
\end{eqnarray}
Using the Fierz rearrangement relation
\begin{eqnarray}
T^{A}_{\alpha\beta}T^{A}_{\rho\lambda}=-\frac{1}{6}\delta_{\alpha\beta}
\delta_{\rho\lambda}+\frac{1}{2}\delta_{\alpha\lambda}\delta_{\rho\beta}\,,
\end{eqnarray}
The relations among the  operators ${\cal O}_{0,8}(\mu)$ and ${\cal O}_{1,2}(\mu)$ can be obtained by
\begin{eqnarray}
{\cal O}_{0}={\cal O}_{1}\,,~~{\cal O}_{8}=-\frac{1}{6}{\cal O}_{1}+\frac{1}{2}{\cal O}_{2}\,.
\end{eqnarray}
Consequently, for the Wilson coefficients, we have
\begin{eqnarray}
C_{0}=C_{1}+C_{2}/3\,,~~C_{8}=2 C_{2}\,,
\end{eqnarray}
where the explicit expression can be found in Ref.~\cite{Qiao:2012hp}.

 In Fig.~\ref{Fig-diagram}, the color-singlet operator contributes to the sub-figure (a) while
the color-octet operator contributes to the sub-figure (b). Here
we  ignored the light meson mass for simplification. Through the numerical calculations, the color-octet operator contribution at leading order is small, which is around 5\% of the color-singlet operator contribution. The color-singlet operator contributions dominate the decay amplitudes, which correspond to the factorizable diagrams. The color-singlet operator matrix element of $B_c^+\to h_c(\chi_{c1} )+\pi^+$ is 
\begin{eqnarray}
&&\langle   h_c(\chi_{c1} ) +\pi^+\vert {\cal O}_{0}\vert B_{c}^{+}\rangle\nonumber\\&=&
 2f_{\pi} A^{h_c(\chi_{c1} )}_{0}(0)m_{h_c(\chi_{c1} )}\varepsilon_{h_c(\chi_{c1} )}^*\cdot q \nonumber\\
&=&  f_{\pi}A^{h_c(\chi_{c1} )}_{0}(0)(m_{B_c}^2-m^2_{h_c(\chi_{c1} )})\,,
\label{pi2}
\end{eqnarray}

\begin{table*}[ht]
\begin{center}
\caption{The exclusive two-body decays of the $B_c$ meson into $P$-wave charmonium and a light meson. The form factors in Tab.~\ref{table1} are employed, the lifetime of the $B_c$ meson is $0.507\times 10^{-12}s$~\cite{Olive:2016xmw}. In NRQCD approach, the relativistic corrections are included in the form factors. The scale is set at the heavy quark mass $m_b$,
and we adopt  $m_b=(4.7\pm0.1)$GeV and $m_c=(1.5\pm0.1)$GeV. The first column uncertainty is from the choice of $m_b$, while the second column uncertainty is from the choice of $m_c$.
}
\begin{tabular}{ccccc}
 \hline\hline
Branching ratios ($10^{-3}$) & NRQCD LO+RC & LFQM~\cite{Wang:2009mi} & QCD SR~\cite{Azizi:2009ny}
   \\
\hline
$\mathcal{B}(B_{c}^{\pm }\to h_{c} +\pi ^{\pm }) $
& $9.73^{+2.75+1.43}_{-2.00-0.90}$ &$0.90$ & $0.002$ \\
$\mathcal{B}(B_{c}^{\pm }\to \chi _{c0} +\pi ^{\pm }) $ & $6.47^{+1.75+0.64}_{-1.26-0.38}$ &$0.53$ & $1.07$ \\
$10^{2}\times \mathcal{B}(B_{c}^{\pm }\to \chi _{c1} +\pi ^{\pm }) $ & $6.38^{+0.77+0.77}_{-1.41-0.72}$ &$4$ & $0.2$ \\
$\mathcal{B}(B_{c}^{\pm }\to \chi _{c2} +\pi ^{\pm }) $ & $4.37^{+1.58+1.58}_{-1.09-1.06}$ &$0.57$ & \\
$10\times \mathcal{B}(B_{c}^{\pm }\to h_{c} +K^{\pm }) $
& $7.42^{+0.21+0.10}_{-0.15-0.06}$ &$0.7$ & $0.002$ \\
$10\times \mathcal{B}(B_{c}^{\pm }\to \chi _{c0} +K^{\pm }) $ & $4.94^{+1.33+0.49}_{-0.96-0.29}$ &$0.4$ & $0.8$ \\
$10^{3}\times \mathcal{B}(B_{c}^{\pm }\to \chi _{c1} +K^{\pm }) $ & $4.87^{+0.59+0.59}_{-1.07-0.55}$ &$3$ & $0.2$ \\
$10\times\mathcal{B}(B_{c}^{\pm }\to \chi _{c2} +K^{\pm }) $ & $3.33^{+1.23+1.21}_{-0.83-0.81}$ &$0.4$ & \\
 \hline
\end{tabular}\label{table2}
\end{center}
\end{table*}

Similarly, the color-singlet operator matrix element of $B_c^+\to \chi_{c0} +\pi^+$ can be written as
\begin{eqnarray}
\langle  \chi_{c0} +\pi^+\vert {\cal O}_{0}\vert B_{c}^{+}\rangle=
f_{\pi} f^{\chi_{c0} }_{0}(0)(m_{B_c}^2-m^2_{\chi_{c0}})\,.&&
\label{pi3}
\end{eqnarray}

And, the color-singlet operator matrix element of $B_c^+\to \chi_{c2}+\pi^+$ is
\begin{eqnarray}
&&\langle  \chi_{c2} +\pi^+\vert {\cal O}_{0}\vert B_{c}^{+}\rangle\nonumber\\&\approx&
 \frac{2}{m_{B_c}}
f_{\pi} A^{\chi_{c2} }_{0}(0)m_{\chi_{c2} }\varepsilon^{*\alpha\beta}_{\chi_{c2}}P_\alpha  P_\beta\nonumber\\
&=&
f_{\pi} A^{\chi_{c2} }_{0}(0)\frac{(m_{B_c}^2-m^2_{\chi_{c2}})^2}{2m_{\chi_{c2}}m_{B_c}}\,.
\label{pi1}
\end{eqnarray}

The decay width  can be written as
\begin{eqnarray}
\Gamma(B_c\to H+ \pi)&=&\frac{|\bold p|}{8\pi
m_{B_c}^2}|{\cal M}(B_c\to H+ \pi)|^2,~~~~
\end{eqnarray}
with the final charmonium momentum  $|\bold p|=(m_{B_c}^2-m^2_{H})/(2m_{B_c})$ in the $B_c$ meson rest frame.

Using the above results of the form factors, we can easily evaluate the branching ratios of  the exclusive two-body decays $B_c \to h_c +\pi $, $B_c \to h_c +K $, $B_c \to \chi_{cJ} +\pi $,  and $B_c \to \chi_{cJ} +K $. The branching ratios are given in Tab.~\ref{table2}.  Compared with the results of LFQM and QCD SR, our results of the  branching ratios are larger. 
From the table, one can easily get the branching ratio of ${\cal B}(B_c^\pm \to \chi_{c0} +\pi^\pm)=(6.47^{+1.75+0.64}_{-1.26-0.38})\times 10^{-3}$. Using the data
$\frac{\sigma(B_c^+)}{\sigma(B^+)}\times {\cal B}(B_c^+\to \chi_{c0} \pi^+)=(9.8_{-3.0}^{+3.4}(stat)\pm0.8(syst))\times 10^{-6}$ observed by the LHCb Collaboration, the cross section ratio $\frac{\sigma(B_c^+)}{\sigma(B^+)}$ can be extracted around $(1.1-2.0)\times 10^{-3}$, which is also useful to study the total cross section of $B_c$ meson at LHC~\cite{Aaij:2014ija}. Considering the current experiment environment and the possible decay events, the channels $B_c^\pm\to \chi_{c0} +K^\pm$  and $B_c^\pm\to h_{c} +\pi^\pm$ have the potential to be reconstructed in addition to $B_c^\pm\to \chi_{c0} +\pi^\pm$.

\section{Conclusion}
 We obtained the relativistic corrections to the form factors of $B_c$ into $P$-wave charmonium at the ${\cal O}(|\bold{k}|^2)$ and ${\cal O}(|\bold{k}'|^2)$level, where
$k$ is a half of reduced  heavy quark relative momentum inside the $B_c$ meson and $k'$ is a half of quark relative momentum inside the $P$-wave charmonium. The heavy quark relative momenta are  small quantities compared with the heavy quark mass. The analytic expressions of the form factors are given in the paper, and we have studied the asymptotic behaviour of the form factors in the heavy bottom quark limit, i.e., $m_b\to \infty$. In this limit, the form factors are not independent and some of them become identical, which reduces the degree of freedom of the form factors.

The values of the form factors in the maximum momentum recoil point with $q^2=0$ are also given. In the weak effective theory, we obtained the decay amplitudes for the decay channels $B_c \to h_c +\pi $, $B_c \to h_c +K $, $B_c \to \chi_{cJ} +\pi $,  and $B_c \to \chi_{cJ} +K $, where the factorizable diagrams dominate the contribution in the heavy bottom quark limit. Our results are consistent with the experimental observation. Using the LHCb data
$\frac{\sigma(B_c^+)}{\sigma(B^+)}\times {\cal B}(B_c^+\to \chi_{c0} \pi^+)=(9.8_{-3.0}^{+3.4}(stat)\pm0.8(syst))\times 10^{-6}$, the cross section ratio $\frac{\sigma(B_c^+)}{\sigma(B^+)}$ can be extracted around $(1.1-2.0)\times 10^{-3}$ at $\sqrt{s}=7-8$TeV. The predictions of the branching ratios of the exclusive two-body $B_c$ decays into $P$-wave charmonium and a light meson can be tested by future LHCb experiments.

\section*{Acknowledgments}
This work was supported in part by the National Natural Science Foundation
of China under Grant No. 11647163 and 11705092, and by Natural Science Foundation of
Jiangsu under Grant No.~BK20171471.

\end{document}